\documentclass[a4paper,11pt,reqno]{amsart}
\pdfoutput=1

\usepackage{amsmath}
\usepackage{amssymb}
\usepackage{color}
\usepackage{tikz}
\usepackage{subfig}
\usepackage{placeins}
\usepackage{booktabs}
\usepackage[margin=0.90in]{geometry}

\title[]{Experimental and numerical investigation of reactive species transport around a small rising bubble}

\author[A.\ Weiner]{Andre Weiner}
\address[]{Mathematical Modeling and Analysis, Technical University of Darmstadt, Germany}
\email[]{weiner@mma.tu-darmstadt.de}
\author[J.\ Timmermann]{Jens Timmermann}
\address[]{Institute of Multiphase Flows, Hamburg University of Technology, Germany}
\email[]{jens.timmermann@tuhh.de}
\author[C.\ Pesci]{Chiara Pesci}
\address[]{Mathematical Modeling and Analysis, Technische \mbox{Universit{\"a}t} Darmstadt, Germany}
\email[]{pesci@mma.tu-darmstadt.de}
\author[J.\ Grewe]{Jana Grewe}
\author[M.\ Hoffmann]{Marko Hoffmann}
\address[]{Institute of Multiphase Flows, Hamburg University of Technology, Germany}
\email[]{marko.hoffmann@tuhh.de}
\author[M.\ Schl\"uter]{Michael Schl\"uter}
\address[]{Institute of Multiphase Flows, Hamburg University of Technology, Germany}
\email[Corresponding author]{michael.schlueter@tuhh.de}
\author[D.\ Bothe]{Dieter Bothe}
\address[]{Mathematical Modeling and Analysis, Technische \mbox{Universit{\"a}t} Darmstadt, Germany}
\email[Corresponding author]{bothe@mma.tu-darmstadt.de}

\date{\today}
\subjclass{flu-dyn}

\begin{document}

\begin{abstract}
  In this article, we present experimental and numerical techniques to investigate the transfer, transport, and reaction of a chemical species in the vicinity of rising bubbles. In the experiment, single oxygen bubbles of diameter $d_b=0.55\dots 0.85~mm$ are released into a measurement cell filled with tap water. The oxygen dissolves and reacts with sulfite to sulfate. Laser-induced fluorescence is used to visualize the oxygen concentration in the bubble wake from which the global mass transfer coefficient can be calculated. The ruthenium-based fluorescent dye seems to be surface active, such that the rise velocity is reduced by up to $50~\%$ compared to the experiment without fluorescent dye and a recirculation zone forms in the bubble wake. To access the local mass transfer at the interface, we perform complementary numerical simulations. Since the fluorescence tracer is essential for the experimental method, the effect of surface contamination is also considered in the simulation. We employ several improvements in the experimental and numerical procedures which allow for a quantitative comparison (locally and globally). Rise velocity and mass transfer coefficient agree within a few percents between experiment, simulation and literature results. Because the fluorescence tracer is frequently used in mass transfer experiments, we discuss its potential surface activity.
\end{abstract}

\maketitle

\section{Introduction}
\label{sec:introduction}
  The mass transfer from a gaseous, dispersed phase into the surrounding liquid phase is a major task within the chemical industry as well as in bio-, food- or environmental engineering. The effective utilization of the gaseous phase with a minimum in energy consumption, optimal contact times and high mass transfer rates has gained more and more importance for process intensification and optimization. Nevertheless, a reliable and exact design of multiphase reactors is one of the unsolved challenges in process engineering since local mass transfer processes in dense bubbly flows cannot be investigated easily. Instead, it is common practice to investigate the behavior of a single bubble and then to apply case-dependent correction factors to account for the effect of bubble-bubble interactions. But also correlations for single fluid particles are mainly empirical or semi-empirical and often ambiguous. Semi-empirical correlations for the Sherwood number ($Sh$) typically have the form of equation \eqref{equ:Brauer} below. The idea behind the equation is as follows: the transport of a passive chemical species around a fluid particle rising in a stagnant liquid may be characterized by the Reynolds ($Re=d_b U_b / \nu$) and Schmidt ($Sc=D/\nu$) number, where $d_b$ is the bubble diameter, $U_b$ is the rise velocity, $\nu$ is the liquid side kinematic viscosity, and $D$ is the molecular diffusivity of the dissolved gas. A simplified analysis based on boundary layer theory leads to a scaling of $Sh \sim Pe^{1/2}$ with the P\'elect number defined as $Pe = Re Sc$ (see for example reference \cite{lochiel1964}). The actual scaling may deviate, for example, if the flow around the bubble is complex (recirculation, vortex shedding), if the bubble is deformed, or if the shape oscillates (wobbling). The simple scaling law also breaks down if the transport is not convection-dominated. For example, as $Re\rightarrow 0$, the Sherwood number of a spherical particle should approach $Sh=2$ and not zero. This theoretical considerations presumably led Brauer \cite{Bra71} to suggest the empirically corrected equation

\begin{equation}
\label{equ:Brauer}
    Sh = 2 + C_1 Re^a Sc^b\ ,
\end{equation}
containing case-specific constants $C_1$, $a$, and $b$. Another critical factor is the presence of surface contamination. As contaminant, we consider any substance in the system which adsorbs at the interface modifying the surface properties, in particular, the surface tension of the gas-liquid interface. Even the presence of tiny amounts of such a substance may significantly modify the mass transfer performance of a system because the interface is partially immobilized. From this point of view, most systems are influenced by contamination. Already the usage of tap water instead of purified water may reduce the rise velocity up to $50~\%$ (see figure 7.3 \cite[p. 172]{clift1978}). Thus, it is not too surprising that experimental results often agree well with the scaling behavior of solid particles $Sh \sim Re^{1/2} Sc^{1/3}$. A frequently cited correlation is that of Fr\"ossling \cite{Fro38}:

\begin{equation}
\label{equ:froessling}
    Sh= 2 + 0.552 Sc^{1/3} Re^{1/2}\ .
\end{equation}
Note that the coefficient preceding the Schmidt number in equation \eqref{equ:froessling} is actually based on experimental results of heat transfer at falling droplets. If the transfer species reacts close to the interface with some bulk component, the (transfer) species boundary layer is depleted and the mass transfer enhanced. Instead of defining correlations for a reactive Sherwood number, it is more common to apply an enhancement factor to the non-reactive value, e.g., $Sh_r = Sh\ E$. A frequently used correlation for $E$ is based on film theory and, for slow reactions, reads as 

\begin{equation}
\label{equ:enhancement_film_theory}
    E = Ha / \mathrm{tanh}(Ha)\ ,
\end{equation}
where the Hatta number is defined as $Ha = \sqrt{kD}/k_L$ for a first-order reaction with the reaction rate $k$ and the non-reactive mass transfer coefficient $k_L$.

Even though experimental and numerical results reflect theoretical scaling laws, there is some ambiguity when it comes to the precise quantitative prediction of mass transfer, or when results differ from the expected scaling behavior. Different measurement or simulation techniques will lead to different correction factors that match the data. The purity of materials used in the experiment or the cleanliness of apertures may significantly affect the obtained results. For example, it is known that rise path and velocity of bubbles rising in water depend strongly on the water's purity. The purity of standard tap water may change on a daily basis in one location, and it certainly varies between different locations (cities). Many times it is not possible or not even wanted to exclude such effects because the ultimate goal is to predict the performance of real-world processes, and, therefore, experiments should resemble their industrial counterparts as closely as possible. On the other hand, experiments also aim to focus on isolated aspects to gain an understanding of the underlying mechanisms. The gap between these two conflicting goals can be reduced by the aid of numerical tools. Simulations are based on mathematical models which aim to describe physical phenomena. When simulation and experiment agree within a few percents with respect to the main performance parameters like rise velocity or mass transfer coefficient, the dominating effects in the experiment can be considered as understood. Furthermore, simulations contain a higher degree of information which allows to comprehend \textit{why} predictions agree or disagree with expectations. To leverage this synergy effect between experiment and simulation, numerical tools have to include all the principal physical phenomena and they have to be able to investigate these effects on all relevant time and length scales.

The present work focuses on a cross-validation between experiments based on the (planar) laser-induced fluorescence (LIF) technique \cite{Bor05, Dan07, Kuc10, Kuc11, Kuc12, Jim14, Hua17} and numerical results based on Arbitrary Eulerian-Lagrangian (ALE) Interface-Tracking \cite{jasTuk2006, tukJas2008, tukJas2012, pesci2015, pesciSPP2017, Pesci2017}. Both techniques allow to record local transport phenomena with high temporal and spatial resolution. We reproduced the experimental setting by K\"uck et al. \cite{Kuc11, Kuc12}, in which relatively small single rising oxygen bubbles were investigated. The bubble diameter ranges between $0.55~mm$ and $0.85~mm$, such that a mostly rectilinear path is attained. The aim is to reduce the complexity by avoiding the additional influence of path instability and shape oscillations. The reaction system is the same as in our previous works, too. The sulfite-sulfate oxidation has several advantages and disadvantages, which are briefly discussed in section \ref{sec:substance_system}. Thanks to developments regarding the frame rate and the laser sheet structure on the experimental side, and a numerical technique that is able to handle surface contamination as well as realistic material properties, we were able to find a promising qualitative and quantitative agreement between experiment, simulation and literature results. The investigation techniques and their improvements are explained in section \ref{sec:methods_investigation}. Results for the rise velocity, the mass transfer coefficient and the oxygen distribution in the bubble wake are presented in section \ref{sec:results}. The latter section also discusses the ruthenium-based fluorescence tracer as a potential surface-active agent (surfactant) and shows how even a small amount of surfactant drastically changes the mass transfer characteristics.

\section{Sulfite-sulfate-reaction}
\label{sec:substance_system}
  Understanding the mixing in gas-liquid contactors is a crucial step for process-intensification. A possible strategy to improve the mixing without additional energy consumption is to exploit the bubble wake dynamics as a local ``mixing device''. To investigate and quantify this strategy, the reaction timescale must be tunable to control precisely where the reaction takes place. The selection of a suitable reaction system is even more challenging for mass transfer investigations based on laser-induced fluorescence techniques, because the transfer species must be fluorescent or quenching. For a gas-liquid interface, a characteristic timescale describing the contact time may be defined as $\tau_{conv} = d_b/U_b$. The timescale of a first-order reaction is simply the inverse of the reaction rate constant $\tau_r = 1/k$. An important dimensionless group defined as the ratio of both timescales is the Damk\"ohler number $Da = \tau_{conv}/\tau_r = kd_b/U_b$. For small values of $Da$, i.e. $Da\ll 1$, the reaction takes place distributed over significant portions of the liquid bulk. The mass transfer enhancement will be negligible, and the influence of local bubble hydrodynamics on the reaction will be small. For small reaction timescales, i.e. $Da \gg 1$, the reaction takes place in the immediate vicinity of the interface. The enhancement will be strong, but it won't be influenced much by the flow in the bubble wake. Additional criteria for a suitable reaction system for method validation may be summarized as follows:

\begin{itemize}
	\item[1)] The physisorption of the transfer species must be known.
	\item[2)] At least one of the educts or products has to be fluorescent or quenching.
	\item[3)] The reaction path should be well-defined and known.
	\item[4)] The reaction timescale should be tunable such that $Da \approx 1$.
\end{itemize}
For the present work, we chose the oxidation of sodium sulfite to sulfate, formally written as

\begin{equation}
\label{eq:mechKueck1}
    2 SO_3^{2-} +  O_2 \xrightarrow{Co^{2+}} 2 SO_4^{2-}\ .
\end{equation}
While the overall reaction is very simple, the detailed reaction mechanism is rather complex due to radical reactions \cite{Bac34}. Actually, there is still some controversy in the literature about the exact reaction steps \cite{She12, Shu12, Lin81, Kor11}. Despite the fact that not all details about the reaction kinetics are clarified yet, we chose this system because it has been widely used in previous studies of reactive mass transfer in different groups \cite{Kuc10, Kuc11, Kuc12, Tim16} and because our aim is to show the progress made in the experimental and numerical techniques.

\section{Methods of investigation}
\label{sec:methods_investigation}

\subsection{Experimental analysis}
\label{sec:experimental_analysis}

\subsubsection{Experimental setup}
\label{sec:experimental_setup}
\begin{figure}[!ht]
	\centering
	\includegraphics[width=0.7\textwidth]{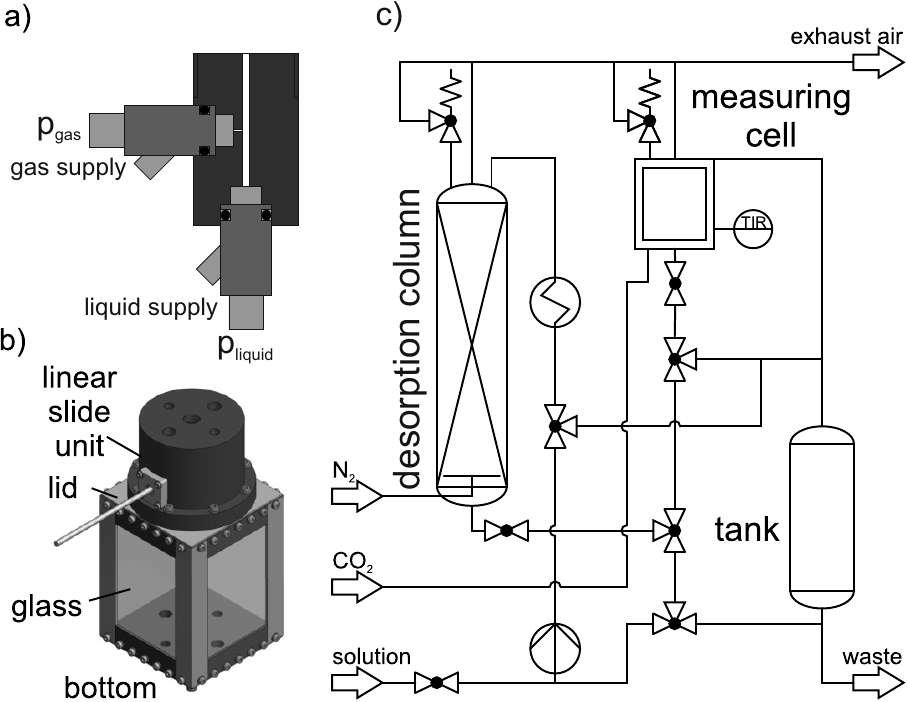}
	\caption{Basic setup for mass transfer investigations with injections valves a), measurement cell b), and flow scheme c). }
	\label{fig:basic-setup}
\end{figure}

  The accurate investigation of mass transfer at small bubbles rising freely on a rectilinear path requires an experimental setup and procedure that allows the generation of bubbles with reproducible shape and trajectories. The generation of rectilinear rising bubbles with a size smaller than 1 mm was already performed by Ohl \cite{Ohl01} and was used previously within the work of K\"uck et al. \cite{Kuc10, Kuc11, Kuc12}. The technique uses two injection valves, one to control the flow rate of the liquid and gaseous flow, and one to adjust the bubble volume by setting the opening period via a function generator; see figure \ref{fig:basic-setup} a). Additionally, the pressure has to be controlled very accurately with precision pressure regulators (range $0.05$ to $2~bar$, adjustable in $0.001~bar$ steps). With this technique bubbles from $0.3$ to $3~mm$ can be generated \cite{Ohl01}. Figure \ref{fig:basic-setup}~c) shows the flow scheme of the basic setup. The setup was used in a slightly different configuration in Timmermann et al. \cite{Tim16}. Heart of the setup is the measuring cell (cross section $150\times 150~mm^2$) which consists of four glass walls with bottom and lid. Both base and top are made of stainless steel to allow different reaction systems. The bubble generator is placed in the center of the bottom part.

\begin{figure}[!ht]
	\centering
	\includegraphics[width=0.5\textwidth]{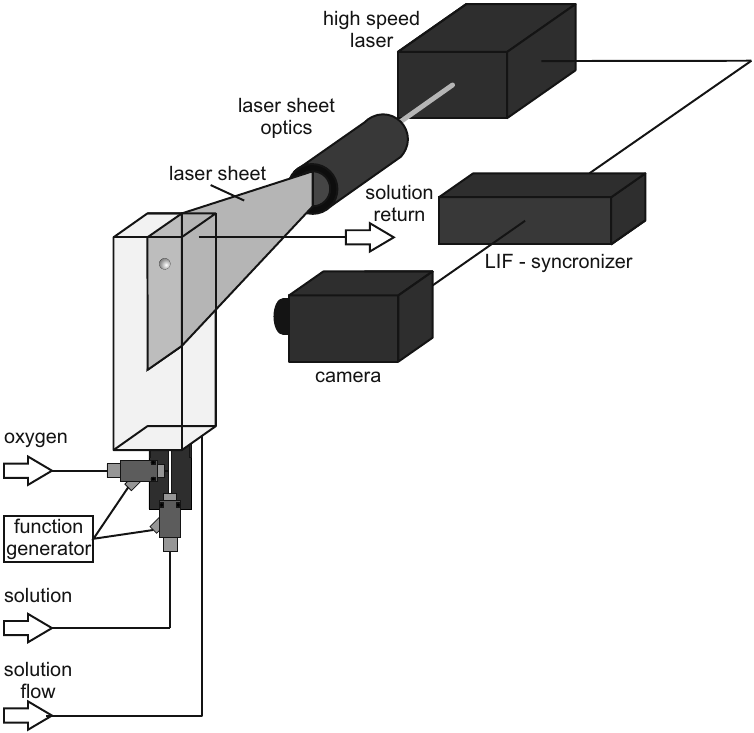}
	\caption{LIF-setup for the investigation of local mass transfer at rectilinear rising oxygen bubbles.}
	\label{fig:LIF-setup}
\end{figure}

For the investigation of local mass transfer at rising gas bubbles, laser-induced fluorescence (LIF) is applied to visualize oxygen concentration fields. The oxygen sensitive dye Dichlorotris-(1,10-phenanthroline)-ruthenium(II)hydrate ($c_{Ru}=30~mg/L$; by Sigma-Aldrich$^{\circledR}$) is used. The fluorescence of this dye shows a dependence on the oxygen concentration and can be described with a Stern-Volmer correlation \cite{Kuc11}. The Stern-Volmer constant and the fluorescent intensity in the absence of oxygen are determined by recording grey level images for several different predefined uniform oxygen concentrations. The oxygen is measured with a PreSens$^{\circledR}$ oxygen sensor. Additionally, the dye solution contains cobalt(II)sulfate ($c_{Co}=16.67~mg/L$) as a catalyst for the sodium sulfite reaction. Excitation of the fluorescent dye is performed with a Nd:YLF laser (wavelength $527~nm$, pulse width $>210~ns$, pulse repetition rate $1~kHz$, Continuum$^{\circledR}$). The laser beam is widened with light sheet optics to obtain a planar laser sheet in the direction of the rectilinear bubble rise. The emitted fluorescence light is recorded with a PCO Dmiax HS2 ($1000~fps$), protected from direct laser radiation by a bandpass filter perpendicular to the laser sheet (ILA\_5150 GmbH, center wavelength $590\pm 2 ~nm$, half-power bandwidth $20\pm 2~nm$, transmission $> 84~\%$). The camera is equipped with an Infinity K2/Distamax objective to achieve high magnifications (field of view: $5.9\times 7.6~mm$). The experimental arrangement for the LIF measurements is shown in figure \ref{fig:LIF-setup}.

To visualize the mass transfer in case of physical and reactive mass transfer, it is necessary to work under oxygen-free conditions. Hence the setup is purged with nitrogen for $30~min$, and simultaneously the dye solution is filled into the desorption column and saturated with nitrogen. The oxygen concentration is monitored during the whole measurement within the desorption column and measuring cell. If the oxygen is depleted and the setup is fully purged with nitrogen, the dye solution is filled into the measuring cell.

\subsection{Image processing}
\label{sec:image_processing}
\begin{figure}[!ht]
	\centering
	\includegraphics[width=0.65\textwidth]{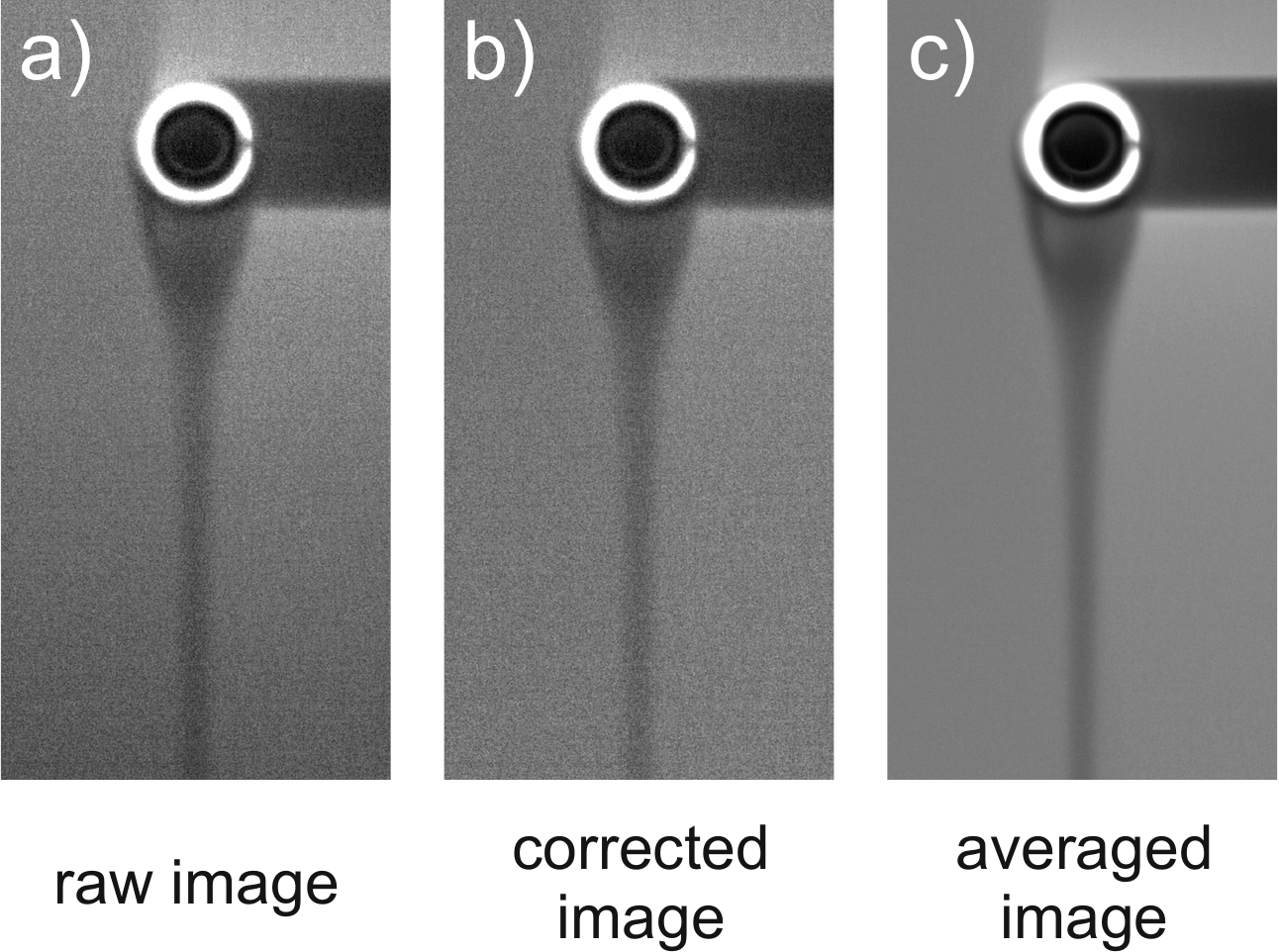}
	\caption{Image processing of raw images.}
	\label{fig:imageprocessing}
\end{figure}

  Due to a non-uniform illumination of the recorded images, a background correction is necessary to obtain reliable concentration field information. The performed correction is based on the work of Dani et al.~\cite{Dan07}. First a background image sequence in oxygen desorbed dye solution is recorded. From this sequence, an averaged image is computed and the recorded raw images, see figure \ref{fig:imageprocessing} a), of the rectilinear rising bubble are divided by the averaged image. Within this corrected image, a nearly uniform background could be obtained; see figure \ref{fig:imageprocessing} b).

In contrast to the experiments of K\"uck et al. \cite{Kuc11, Kuc12},  high-speed recordings of the bubble rise were performed to obtain averaged images with small noise information. Therefore the background corrected images were automatically processed with edge detection in Matlab$^{\circledR}$ to detect the bubble center on each image based on the reflection ring inside the bubble. The found bubble center was used to define a region of interest (ROI)-frame of $400\times 800$ pixels such that the bubble center is always at a position of $200$ pixels in $x$- and $150$ pixels in $y$-direction. Afterwards, the ROI images were averaged to obtain an image with reduced noise level; see figure \ref{fig:imageprocessing} c). Nevertheless, sub-pixel shifts of the bubble center could not be compensated with this technique, so that a small blur of the bubble and its wake structure also occurs.

\subsection{Numerical analysis}
\label{sec:numerical_analysis}

\subsubsection{Numerical setup}
\label{sec:numerical_setup}
  We used an enhanced version of the \textit{bubbleInterTrackFoam} \cite{jasTuk2006,tukJas2008,tukJas2012} solver contained in version 3.1 of foam-extend\footnote{https://sourceforge.net/projects/foam-extend/} (community-driven version of OpenFOAM\textsuperscript{\textregistered}) to perform numerical simulations of clean and contaminated bubbles. The interface-tracking method is complex and computationally expensive, but currently also the most accurate approach to simulate multiphase flows of single fluid particles. One momentum equation is solved for each bulk phase, and the exchange of momentum between the phases occurs via a direct implementation of the jump conditions at the interface. The exact fulfillment of the jump conditions is a major advantage over all methods solving a one-field formulation of the momentum equation; e.g. volume-of-fluid, level-set, front-tracking approaches. Furthermore, the explicit representation of the interface by the computational grid allows to solve surface transport equations. On the other hand, interface tracking solvers are expensive (iterative inter-phase coupling, mesh motion) and less robust than other approaches.

The main modifications are a \emph{sorption-library} \cite{pesci2015, pesciSPP2017} to simulate the effect of soluble surfactant and a \emph{subgrid-scale model} to approximate the convection-dominated transport of surfactant in the boundary layer forming at the bubble interface \cite{Pesci2017}. The sorption-library contains models for fast (diffusion-controlled) and slow (kinetically controlled) sorption as well as corresponding models for the sorption-isotherm. One of the main challenges in this study is the presence of surface-active agents (surfactants) in the experimental system, which alter the properties of the gas-liquid interface. The contaminant source is very likely related to the addition of the fluorescent dye, as will be discussed in section \ref{sec:surface_contamination}. From the experimental results of the bubble rise velocity presented in section \ref{sec:rise_velocity}, it can be assumed that there is at least one surfactant present in the system. We do not know the exact source of the contamination, however, there is one general observation to our advantage: if the surface contamination is sufficiently high, the bubble reaches a steady velocity which is, over a wide range of surfactant concentrations, independent from the exact concentration value in the liquid bulk \cite{Pesci2017}. The bubble dynamics in the initial transient stage may vary from case to case, but here we only consider the steady-state regime. Even though the steady-state rise velocity may not depend on the surfactant concentration, it could still depend on surfactant-specific material parameters like the molecular diffusivity of the surfactant in the liquid bulk and on the interface, or the Langmuir constant. Therefore, we tested the behavior under different bulk concentrations of two commonly used surfactants, namely Triton X-100 and C\textsubscript{12}DMPO. In the tested parameter range, the rise velocity never changed more than $\pm 3~\%$ around the mean of all test-cases. This observation is supported by the fact that in most of the literature concerned with the rise of single bubbles in a contaminated liquid (see for example \cite{Tom98, Alv05, Pesci2017}) the hydrodynamic drag is comparable to the one of a rising solid particle. No strong dependency on the specific surfactant is observed or included in correlations for the drag coefficient. Of course, these assumptions may not hold true in general, but the results presented in the sections hereafter suggest that, at least for the present study, these arguments seem to be justified. For completeness, the exact surfactant properties used in the simulations of contaminated bubbles are reported in table \ref{tab:surfactant_properties}. The initial surfactant bulk concentration was set to $c_0=0.05~mol/m^3$, a very high value which causes the bubble to reach quickly a ``fully contaminated'' state (a state in which the Marangoni forces remain approximately constant). For a more comprehensive discussion of the properties in table \ref{tab:surfactant_properties} and the sorption model, the reader is referred to the sections 2-4 in \cite{Pesci2017}.

\begin{table}[ht]
    \caption{Surfactant (C\textsubscript{12}DMPO) properties, fast Langmuir adsorption model parameters. $c^\Sigma_\infty$ - saturated surface concentration, $a_L$ - Langmuir constant, $D$ - molecular diffusivity in the liquid bulk, $D^{\Sigma}$ - molecular diffusivity on the interface, $T$ - temperature}
    \label{tab:surfactant_properties}
    \centering
    \vspace{0.2cm}
    \begin{tabular}{ccccc}
        \toprule
        $c^{\Sigma}_{\infty}$ in $mol/m^2$ & $a_L$ in $mol/m^3$ &
        $D$ in $m^2/s$ & $D^{\Sigma}$ in $m^2/s$ & $T$ in $K$\\
        \midrule
         $4.17 \cdot 10^{-6}$ & $4.85 \cdot 10^{-3}$ & $5 \cdot 10^{-10}$ & $5 \cdot 10^{-7}$ & $296$\\
        \bottomrule
    \end{tabular}
\end{table}

Further simulation parameters are the bulk phase properties of liquid (water) and gas (oxygen) compiled in table \ref{tab:bulk_properties}. Note that in the numerical model the dissolved oxygen is treated as a passive scalar meaning that it does not influence the liquid properties. Moreover, the volume change of the bubble due to the oxygen is neglected. The oxygen concentration fields presented in later chapters result from the solution of a convection-diffusion equation in the liquid phase. A constant value of $c\vert_{\Sigma,O_2} = 1.331~mol/m^3$ (the saturation concentration of oxygen in water) was set as a boundary condition at the interface. The molecular diffusivity of oxygen $D_{O_2} = 2.0\cdot 10^{-9}~m^2/s$ is by a factor of four larger than that of the surfactant, however, the concentration boundary layer is still extremely thin such that a subgrid-scale model is necessary to approximate the oxygen transfer accurately. The basic modeling idea is unchanged compared to previous works \cite{Weiner2017, Pesci2017}. The analytical solution of a substitute problem is used to improve the numerical solution of the species transport equation close to the interface, where the computational mesh is not fine enough to resolve the steep concentration profile. This approach was used previously for physical species transfer \cite{Weiner2017} and for the sorption of surfactant \cite{Pesci2017}. To handle also species transfer with a first-order chemical reaction, we derived modified model equations based on a ``reactive'' version of the substitute problem, which can be found in \ref{sec:reactive_sgs_model}.

\begin{table}[ht]
    \caption{Hydrodynamic material properties for the liquid ($+$) and gas ($-$) phase. $\rho$ - density, $\mu$ - dynamic viscosity, $\sigma$ - surface tension (clean).}
    \label{tab:bulk_properties}
    \centering
    \vspace{0.2cm}
    \begin{tabular}{ccccc}
        \toprule
        $\rho^+$ in $kg/m^3$ & $\rho^-$ in $kg/m^3$ & $\mu^+$ in $kg/(ms)$ & $\mu^-$ in $kg/(ms)$ & $\sigma$ in $N/m$\\
        \midrule
         $1000$ & $1.331$ & $1\cdot 10^{-3}$ & $2 \cdot 10^{-5}$ & $0.0724$\\
        \bottomrule
    \end{tabular}
\end{table}

\begin{figure}[ht]
    \centering
    \subfloat[][\emph{Full domain}.]
    {\includegraphics[width=.45\textwidth]{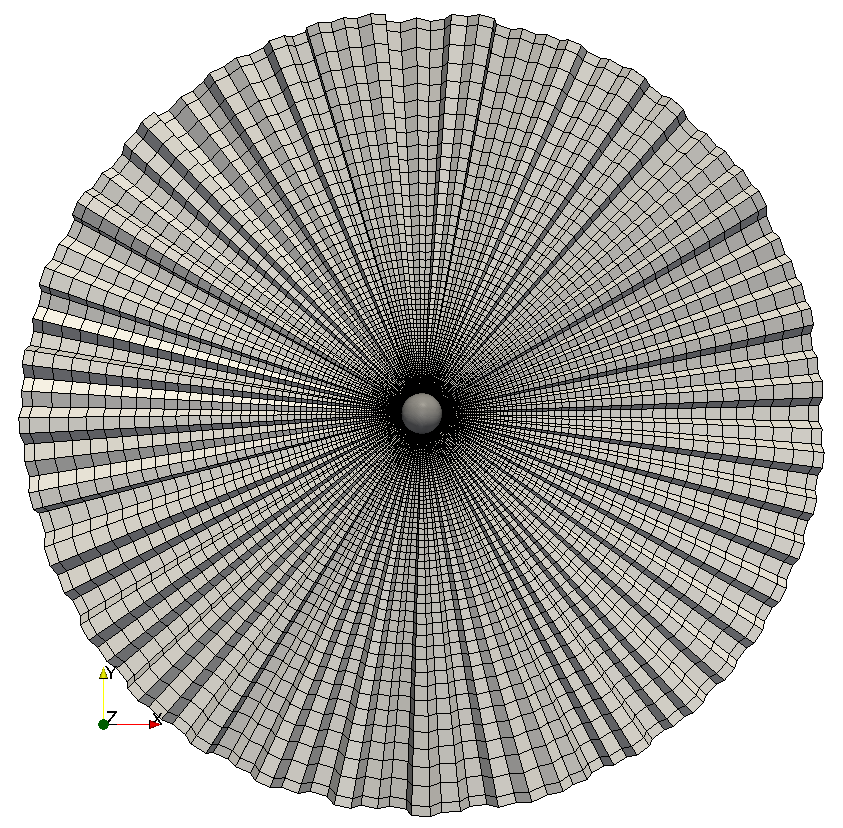}\label{fig:v8a}} \quad
    \subfloat[][\emph{Enlarged view of the bubble region}.]
    {\includegraphics[width=.45\textwidth]{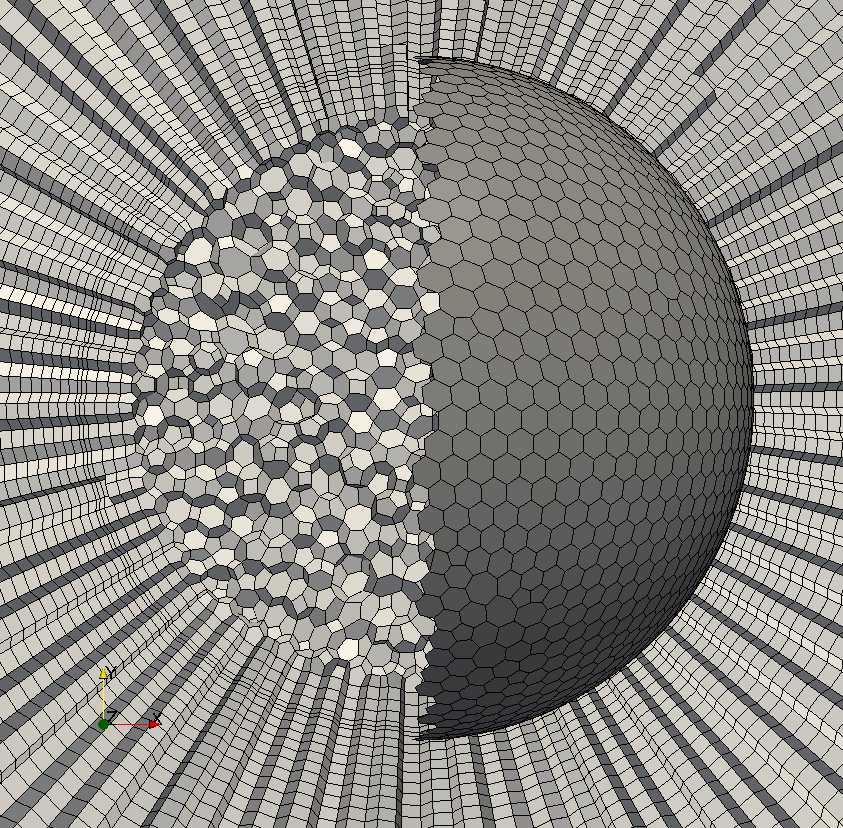}\label{fig:v8b}}
    \caption{3D computational domain for a rising bubble. Inner, outer and surface meshes (the surface mesh is depicted in a darker grey).}
    \label{fig:computational_mesh}
\end{figure}

Due to the enormous computational costs of the numerical method, we simulated only three different bubble diameters $d_b=0.70$/$0.75$/$0.80~mm$. After the bubbles reached a steady-state, velocity field and bubble shape were frozen, and the reactive mass transfer was solved. This procedure allows to investigate a larger range of reaction rates since the solution of a single transport equation requires significantly less computational effort. The meshes have about $220$/$241$/$278$ thousand bulk and $2850$/$3124$/$3607$ interface cells for the bubble diameters $d_b=0.70$/$0.75$/$0.80~mm$, respectively. Figure \ref{fig:computational_mesh} shows, as an example, the mesh for $d_b=0.75~mm$. We also ran simulations on meshes with approximately twice as many bulk cells and found the rise velocity to vary by less than $0.2~\%$ for both the clean and contaminated case.

\subsubsection{Post-Processing of concentration fields}
\label{sec:post_processing}
\begin{figure}[ht]
	\centering
    \includegraphics[width=0.3\textwidth]{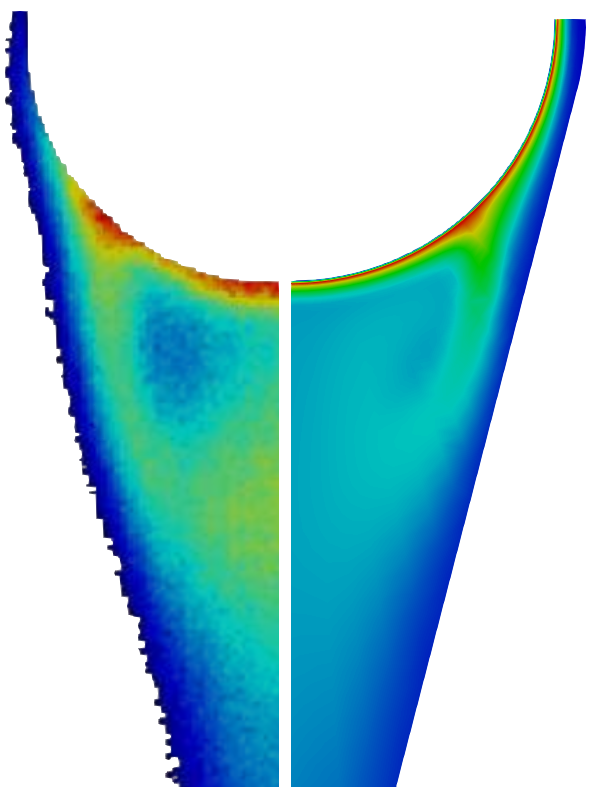}
	\caption{Qualitative comparison of oxygen distribution in the bubble wake: experiment (left half) and cut-plane through computational domain (right half). In the experimental recording, areas with strong reflections close to the gas-liquid interface have been removed; see section \ref{sec:wake_concentration}.}
	\label{fig:experiment_vs_simulation_conc}
\end{figure}

  The measuring principle of fluorescence quenching is based on the effect that oxygen molecules absorb energy of the excited fluorescence tracer. As a result, the fluorescence intensity in regions rich of oxygen is reduced. Based on a calibration curve, gray shades in the recorded camera images can be related to quantitative oxygen concentration values. The laser sheet excites the tracer molecules in a plane parallel to the rise direction and therefore allows to visualize the concentration field in the bubble wake. A similar result may be extracted from the simulation by ``slicing'' through the computational domain. Mathematically, such a slice or plane is a 2D object with zero thickness. For the comparison of local fields between experiment and simulation, it is essential to understand how far the laser-sheet represents such an object. Figure \ref{fig:experiment_vs_simulation_conc} is a preview on the results presented in section \ref{sec:wake_concentration}. The shape of the wake and the angle at which the flow detaches from the interface agree very well between experiment and simulation. However, there are some areas of higher concentration which appear in the experimental recording but not in the simulation. A thorough discussion follows in section \ref{sec:wake_concentration}.

The laser-sheet has a width of approximately $0.5~mm$ in the region where it intersects the rising bubble. When considering bubble diameters of less than $1.0~mm$ this thickness is not negligible. The numerical post-processing step introduced below is a simple attempt to investigate the influence of the finite laser-sheet width. It is conceivable that the measured concentration field is a depth-averaged version of the 3D field in the bubble wake, where the integration/averaging length corresponds to the width of the laser-sheet. Note that this will not alter the absolute concentration value since the calibration curve already contains the  same effect. It is more likely that the observed concentration field might become blurry or smeared out. To verify or falsify this assumption, we emulate the same effect in the post-processing of the numerical results.

\begin{figure}[ht]
	\centering
    \includegraphics[width=0.6\textwidth]{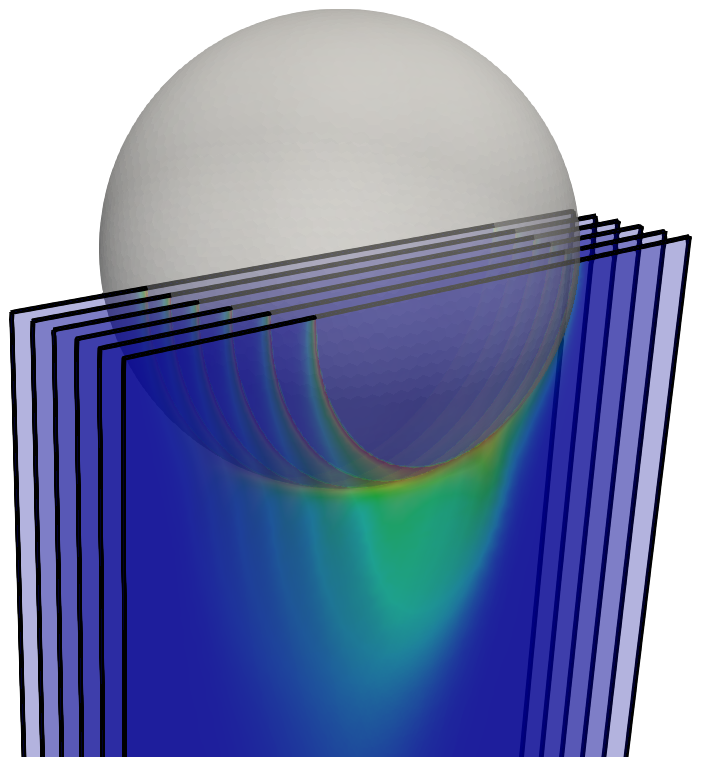}
	\caption{Example slices to extract concentration fields.}
	\label{fig:extracted_slices}
\end{figure}

Figure \ref{fig:extracted_slices} depicts how (triangulated) slices of the concentration field are extracted. This intermediate step is necessary because the computational mesh does not have a fixed underlying structure (like a Cartesian mesh) such that it is not straight-forward to integrate or average the concentration field in a given cuboid. With the extracted slices it is possible to reconstruct the field in the volume in between. A simple linear profile between two $x$-$y$-plane at $z_{i-1}$ and $z_i$ leads to the formula

\begin{equation}
\label{equ:c_sector}
    c(x,y,z) = c(x,y,z_{i-1}) + \frac{c(x,y,z_{i})-c(x,y,z_{i-1})}{z_i-z_{i-1}} (z-z_{i-1})\ ,
\end{equation}
where $z$ may be thought of as a coordinate perpendicular to the laser-sheet. The average over $N_{sec}$ sections can then be computed using the trapezoidal rule

\begin{equation}
\label{equ:c_sector_average}
    \left\langle c \right\rangle_z =
    \sum\limits_{i=1}^{N_{sec}} \left[ \frac{c(z_i)-c(z_{i-1})}{2} \left( z_i-z_{i-1} \right) \right]
    /\sum\limits_{i=1}^{N_{sec}}\left( z_i-z_{i-1} \right) ,
\end{equation}
where $z_{i-1}$ and $z_{i}$ are positioned in the planes at the start and end of each section, respectively. The variables $x$ and $y$ in equation \eqref{equ:c_sector_average} were dropped for simplicity. For an equidistant spacing between the planes, the formula reduces to

\begin{equation}
\label{equ:c_sec_average_simple}
    \left\langle c \right\rangle_z =
    \sum\limits_{i=1}^{N_{sec}} \left[ c(z_i)-c(z_{i-1})\right]
    /\left( 2N_{sec} \right) \ .
\end{equation}

\section{Results}
\label{sec:results}

\subsection{Rise velocity}
\label{sec:rise_velocity}
\begin{figure}[ht]
	\centering
    \includegraphics[width=0.9\textwidth]{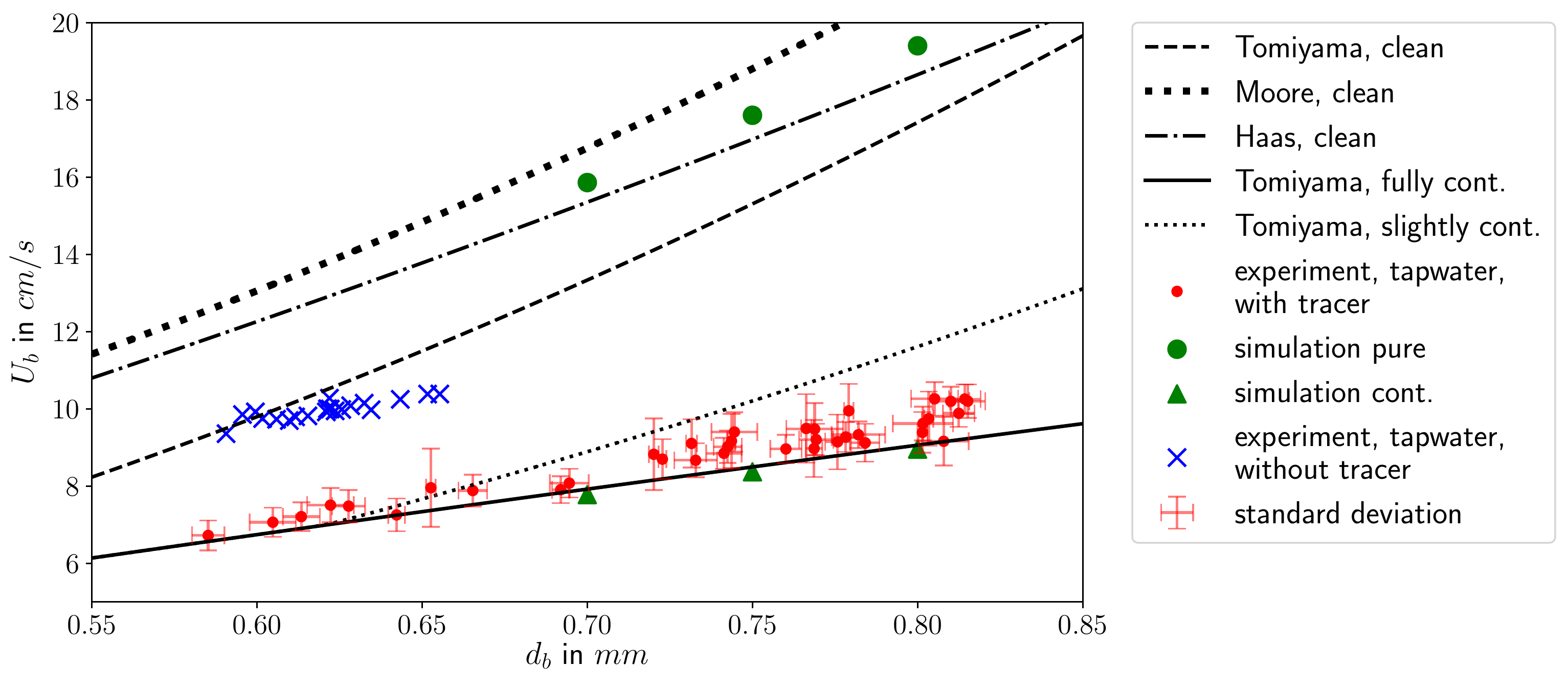}
	\caption{Bubble rise velocity according to simulations, experiments, and correlations. The reference velocities (depicted as lines) were computed based on correlations for the drag coefficient $c_D$. Table \ref{sec:drag_correlations} contains an overview of the different sources, their basis, and their intended usage (clean/contaminated bubbles or solid particles).}
	\label{fig:risevelocity}
\end{figure}

  As mentioned above, the fluorescence tracer seems to reduce the mobility of the bubble surface. Figure \ref{fig:risevelocity} shows the rise velocities obtained from our experiments and simulations compared to reference values from the literature. Relationships for the rise velocity are usually expressed in dimensionless form using the drag coefficient $c_D$. Equating drag and buoyancy force allows to compute a steady-state velocity; see, e.g. equation (5) in \cite{Tom98}):

\begin{equation}
\label{equ:balance_drag_buoyancy}
  U_b^2 = \frac{4(\rho_l-\rho_g)gd_b}{3c_D\rho_l}\ .
\end{equation}
In general, $U_b$ cannot be calculated directly from expression \eqref{equ:balance_drag_buoyancy}, unless a simple correlation for $c_D$ is used\footnote{For example, Levich's $c_D = 48/Re$ leads to an explicit formula. In this work, we applied a bi-section algorithm to compute the reference rise velocity iteratively.}.

Figure \ref{fig:risevelocity} contains several reference lines for the rise velocity which are based on experimental or numerical results of other authors. For clean bubbles, there is usually a tendency that theoretical and numerical results predict higher terminal velocities than experiments of clean water systems. This difference becomes smaller if particular attention is paid in the experiments to reduce any additional influence like contamination or multi-component mass transfer. Typical measures are degassing and water purification. It is therefore not surprising that the numerical predictions are close to the correlations by Haas et al. (based on simulations) and Moore et al. (based on boundary layer theory). Instead, the experimentally obtained rise velocities in tap water without the fluorescent dye are close to Tomiyama's reference curve (based on experiments). Note that the results without the fluorescence tracer stem from preliminary tests conducted before the mass transfer experiments. For these cases, rise velocity and diameter are estimated based on edge-detection such that the error margins were small and are therefore not depicted in the plot. On the other hand, in the mass transfer experiments based on LIF, there is only a low gray level difference between the bubble surface and the surrounding liquid. Furthermore, the gray level varies due to the bubble's wake and shadow. Therefore, it is not possible to apply the edge-detection algorithm. Thus, for these cases, the bubble is assumed to be spherical, and velocity and size are determined based on the bubble's shadow; see figure \ref{fig:greylevelimage}. This measurement principle is less accurate and more noisy; see table \ref{tab:relative_std_ub_d} for a quantification.

\begin{table}[ht]
	\centering
	\caption{Minimum, maximum and mean relative standard deviation for the measured $U_b$ and $d_b$.}
	\vspace{0.2cm}
	\renewcommand{\arraystretch}{1.3}
	\begin{tabular}{cccc}
		\toprule
		 & $\mathrm{min}(\sigma_{\varphi}/\varphi)$ & $\mathrm{max}(\sigma_{\varphi}/\varphi)$ & $\mathrm{mean}(\sigma_{\varphi}/\varphi)$ \\
		\midrule
        $\varphi = U_b$ & $0.0311$ & $0.1270$ & $0.0577$ \\
        $\varphi = d_b$ & $0.0019$ & $0.02218$ & $0.0092$ \\
		\bottomrule
	\end{tabular}
	\label{tab:relative_std_ub_d}
\end{table}

\begin{figure}[ht]
	\centering
    \includegraphics[width=0.75\textwidth]{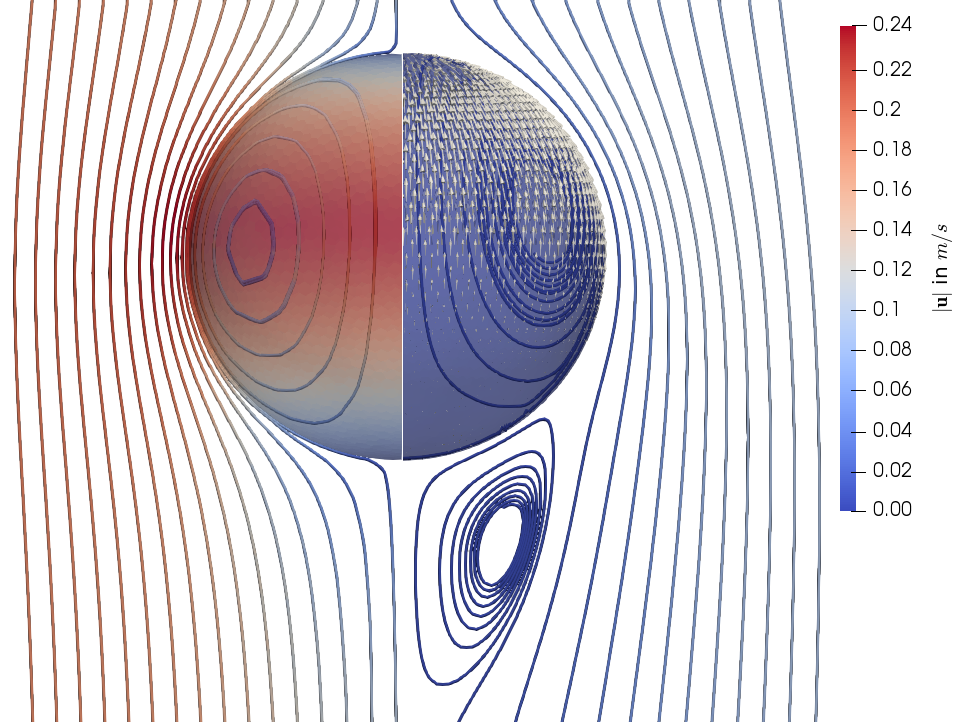}
	\caption{Streamlines around the bubble in a moving reference frame with (right half) and without (left half) surface active agent. The vectors on the interface of the contaminated bubbles represent the local Marangoni forces. Interface and streamlines are colored by the magnitude of the velocity vector.}
	\label{fig:streamlines}
\end{figure}

After adding the fluorescent dye to the system, the observed terminal velocities are significantly reduced and agree well with the reference for fully contaminated systems. The numerically predicted rise velocities for contaminated bubbles are close to the reference curve for solid particles (or fully contaminated bubbles), too. Figure \ref{fig:streamlines} shows the internal and external fluid flow predicted by the simulations. The interface velocity is significantly reduced and almost stagnant (in a reference frame moving with the bubble center). Furthermore, a recirculation zone forms in the bubble wake. The local Marangoni forces associated with the fully contaminated state are distributed over the entire upper bubble hemisphere. A more detailed discussion of the forces acting on the bubble can be found in \cite{Pesci2017}.

\subsection{Oxygen concentration in the wake}
\label{sec:wake_concentration}
\begin{figure}[ht]
	\centering
	\includegraphics[width=0.75\textwidth]{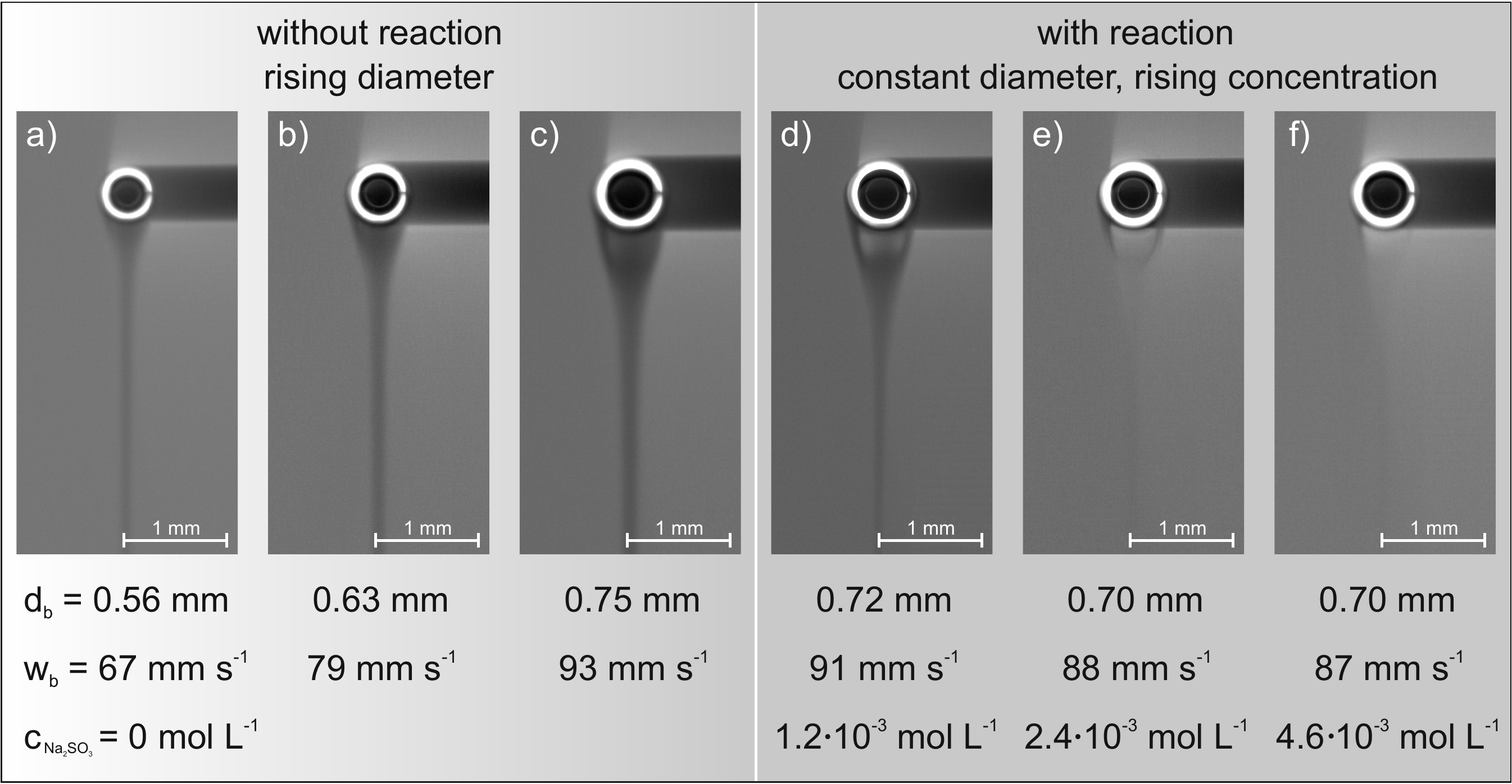}
	\caption{Influence of bubble size and sulfite concentration on the oxygen concentration field in the wake (experimental data).}
	\label{fig:greylevelimage}
\end{figure}

  Figure \ref{fig:greylevelimage} shows the averaged ROI-images in the experiment for several diameters with and without superimposed sodium sulfite reaction. With increasing diameter, the region rich of oxygen becomes broader and longer, as previously reported by Wasowski and Bla\ss \ \cite{Was87}. The observed width of the wake structure indicates the presence of a stagnant ring vortex below the bubble, which is in agreement with the streamlines depicted in figure \ref{fig:streamlines}. When sodium sulfite is added to the experiment, the reaction consumes oxygen. Note that the sulfite concentration is at least one order of magnitude higher than that of oxygen, which is why we discuss the addition of sulfite like an increase of the reaction rate in a pseudo-first-order reaction. For the slowest reaction investigated, a small oxygen-drained region in the wake can be observed, which corresponds to the core of the ring-vortex. The transport of additional oxygen into the stagnant recirculation zone happens mainly due to diffusion and is therefore rather slow, such that the oxygen is completely consumed by the reaction. Moreover, the oxygen distribution is more confined compared to the non-reactive case. With a further increase in the reaction rate or the sulfite concentration, respectively, most of the oxygen is already consumed close to the gas-liquid interface. Note that this observation is very similar to the results of K\"uck et al. \cite{Kuc12}, where for a constant sulfite concentration the reaction rate was modified via the water temperature.

\begin{figure}[ht]
	\centering
	\includegraphics[width=0.75\textwidth]{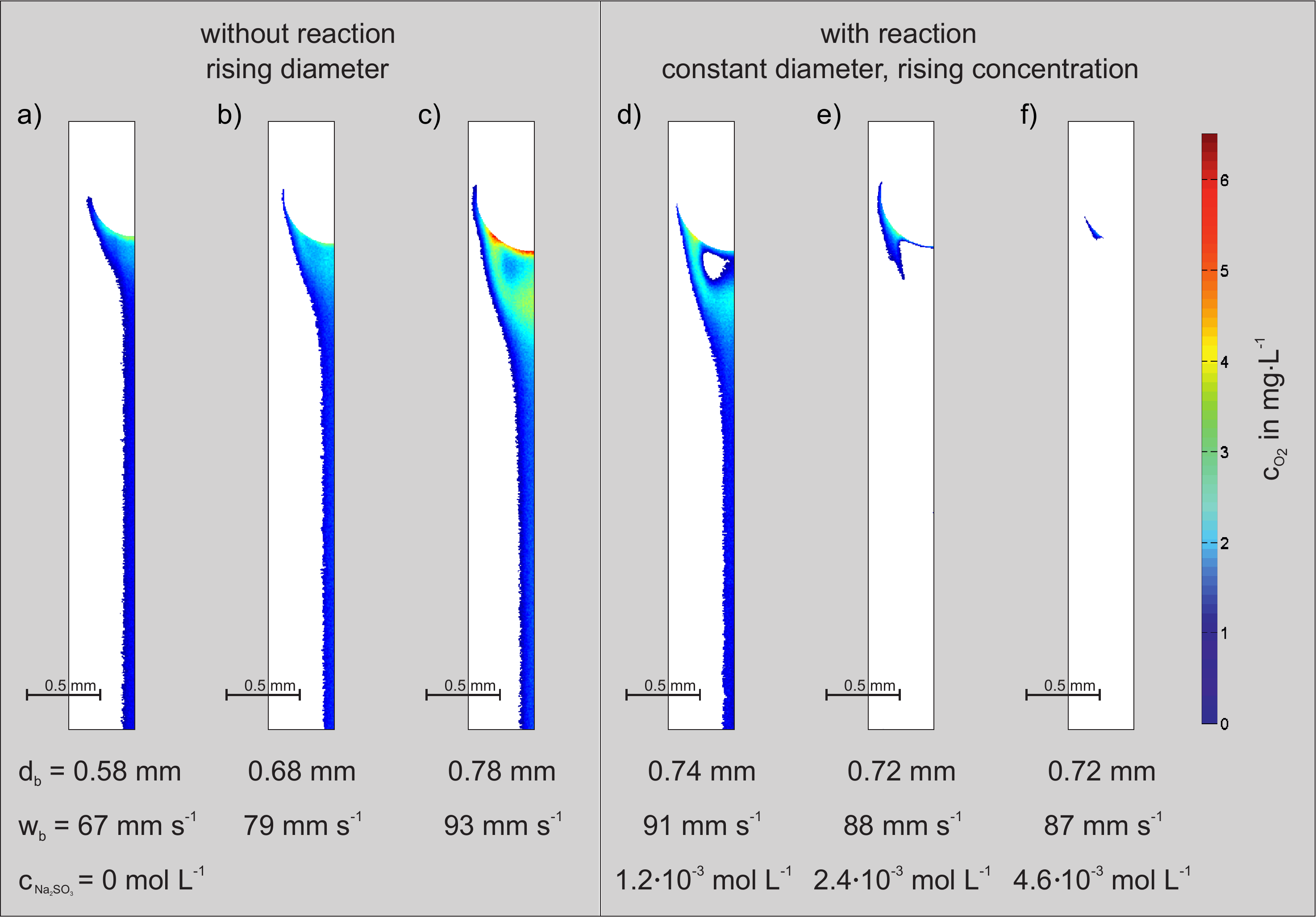}
	\caption{Bubble size and sulfite concentration influence on the oxygen concentration field as pseudo color image (experimental data).}
	\label{fig:pseudocolorimage}
\end{figure}

To determine the mass transfer coefficient and also to have a better visual comparison to the numerical results, the gray-level images were evaluated based on the Stern-Volmer relationship. Figure \ref{fig:pseudocolorimage}  shows the resulting quantitative oxygen distributions. Note that concentrations below $0.1~mg/L$ are depicted white. This threshold also removes the areas with strong reflections close to the gas-liquid interface, and therefore, parts of the boundary layer where the oxygen concentration is presumably higher than the lower threshold. Additionally, the images are cropped to a size of $80\times 783$ pixels.

\begin{figure}[ht]
	\centering
	\includegraphics[width=0.75\textwidth]{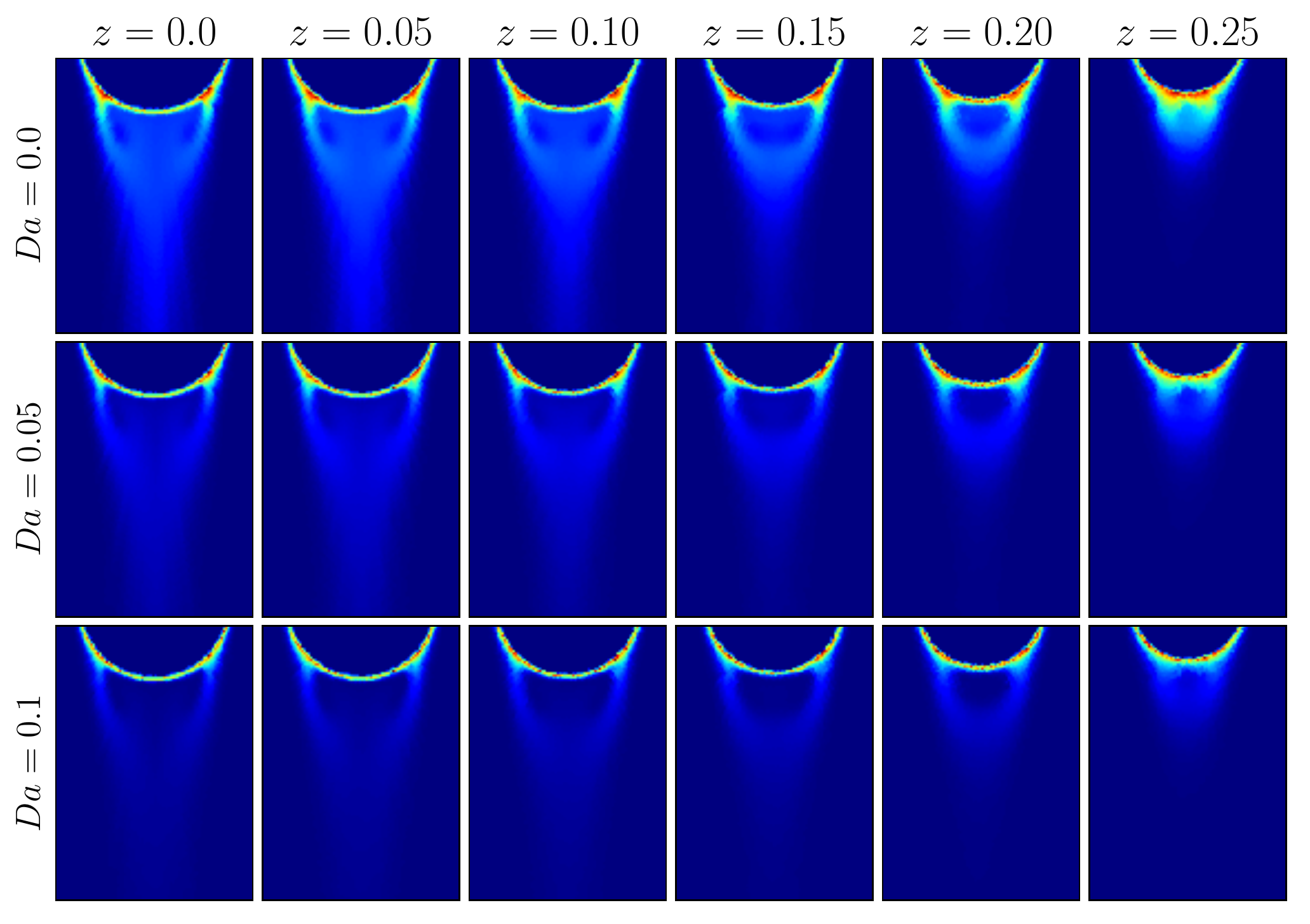}
	\caption{Slices extracted at different distances $z$ (in $mm$) from the bubble center (numerical data).}
	\label{fig:wake_slices}
\end{figure}

The pseudo-color images in figure \ref{fig:pseudocolorimage} reveal more details than their gray counterparts. Taking a close look at the sub-figures c) and d) leads to the following observation: there are two unexpected regions of higher concentration. The first one is close to the interface in proximity of the bubble south-pole. The second region is farther downstream before the recirculation zone ends and the wake becomes closed again. Based on the simulations, one would expect the highest concentration to be measured close to the ring where the flow detaches from the interface since the oxygen transferred into the boundary layer accumulates there. With increasing distance from the interface, the measured concentration of the transfer species (oxygen) should be always lower than close to the interface since there is no mechanism to build up new concentration maxima\footnote{Assuming that the transport of oxygen can be described by a convection-diffusion equation.}. Molecular diffusion or chemical reactions\footnote{Assuming that oxygen is consumed in the reactions.} can only lead to a depletion of oxygen. A possible explanation for the observed experimental concentration distribution will be given hereafter.
As mentioned in section \ref{sec:post_processing}, we assume that the finite width of the laser sheet influences the measured concentration field. Figure \ref{fig:wake_slices} shows slices through the numerically obtained concentration fields extracted at different distances from the bubble center. Looking at the first column, the oxygen distribution is very similar to the experimental result, reflecting the ring-vortex. Because of the immobility of the interface, the contact time between fluid elements and the interface is significantly increased. Even for low Damk\"ohler numbers like $Da=0.1$, the oxygen in the wake is quickly consumed (the characteristic reaction timescale is $10$ times larger than the hydrodynamic timescale). For the last column in figure \ref{fig:wake_slices}, it is much harder to comprehend the concentration field. What we see is more like a slice through the boundary layer than through the wake.

\begin{figure}[ht]
	\centering
	\includegraphics[width=0.70\textwidth]{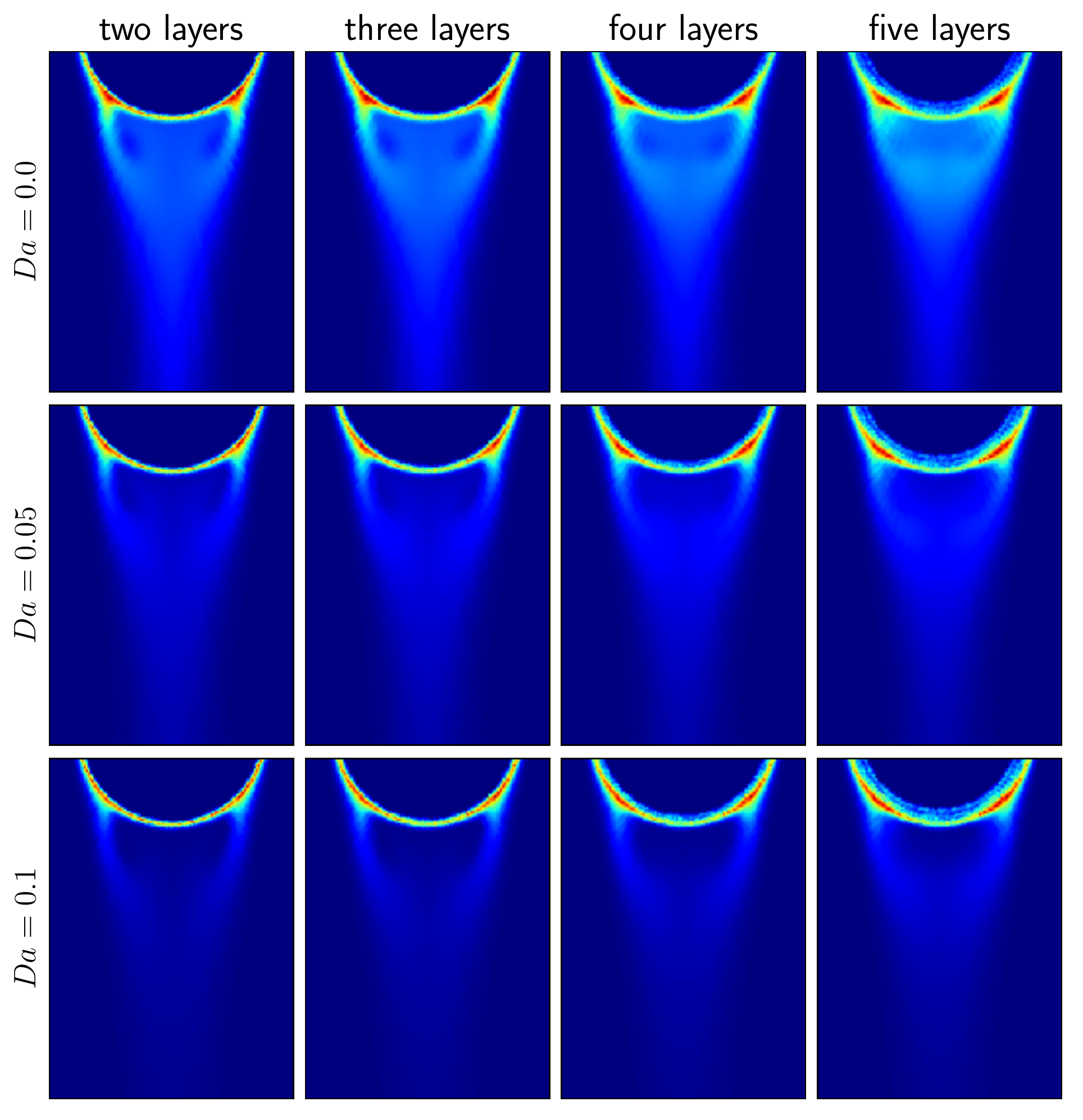}
	\caption{Averaged concentration fields emulating a variation the of laser sheet thickness (numerical data).}
	\label{fig:wake_slices_averaged}
\end{figure}

Figure \ref{fig:wake_slices_averaged} shows the oxygen distribution in the wake obtained if the post-processing procedure described in section \ref{sec:post_processing} is applied. The images in each row from left to right present the averaged concentration field for an increasing width of the laser sheet (averaging volume). For example, images in the first column ``two layers'' correspond to an average over the volume between the first and second slice in figure \ref{fig:wake_slices}, or to a laser sheet thickness of $2\cdot0.05~mm = 0.1~mm$ (the factor $2$ results because we assume symmetry with respect to the bubble center). The image in the last column corresponds to the laser sheet thickness in the experiment. Interestingly, the averaging leads to a region of higher concentration at the end of the recirculation zone as observed in the experiments. Another effect is that the oxygen field close to the interface is blurred, but this area is usually not visible in LIF experiments because of reflections. Possibly one could find even better agreement with the experimental results by varying the averaging volume. For example, averaging only over the volume between the last three or four columns in figure \ref{fig:wake_slices} would presumably lead to a region of high concentration close to the bubble south pole. Further aspects which we have not considered here are that (i) in an experiment it is highly unlikely that the laser sheet insects the bubble perfectly centered, and that (ii) the bubble path always undergoes small oscillations. These effects could further influence the observed results.

\subsection{Local and global mass transfer}
\label{sec:mass_transfer}
\begin{figure}[ht]
	\centering
    \includegraphics[width=0.35\textwidth]{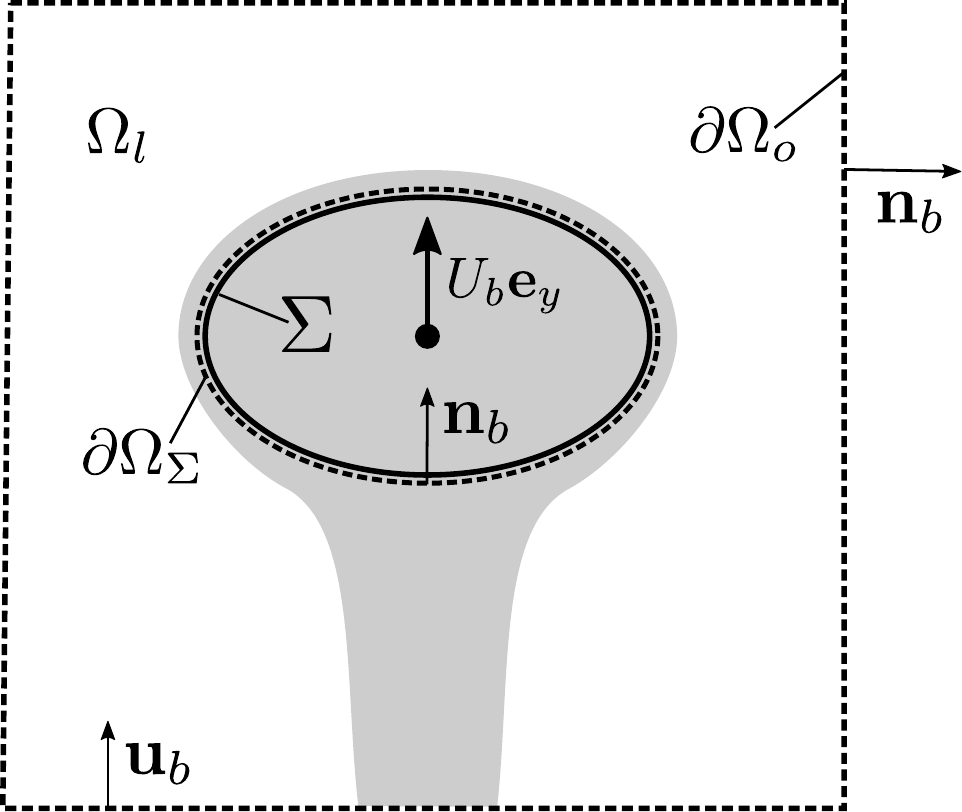}
    \caption{Control volume for the mass balance of oxygen. The vector $\mathbf{e}_y$ denotes the unit normal vector in rise direction.}
	\label{fig:domain_kl}
\end{figure}

  To determine the mass transfer coefficient $k_L$, we follow the procedure of K\"uck et al. \cite{Kuc10, Kuc11, Kuc12}, and Jimenez et al. \cite{Jim14}, but we would like to provide a more detailed derivation which clarifies assumptions and limitations. The global liquid-side mass transfer coefficient $k_L$ is defined as the molar flux $\dot{N}_\Sigma$ over the gas-liquid interface $\Sigma$ per transfer area $A_\Sigma$ and macroscopic concentration difference $\Delta c_{O_2}$, i.e.,

\begin{equation}
\label{equ:massflowrate5}
    k_L = \frac{\left\vert\dot{N_\Sigma}\right\vert}{A_\Sigma \Delta c_{O_2}}\ .
\end{equation}
The concentration difference $\Delta c_{O_2} = {c_{{O_2}}^*}- c_{{O_2},\infty}$ is defined as the difference between the saturation concentration of oxygen at the liquid side of the interface and the bulk concentration far away from the bubble (here we assume $c_{{O_2},\infty} = 0~mol/m^3$). Assuming a steady, spherical bubble shape, the effective mass transfer area is $A_\Sigma = \pi{d_b}^2$. The only missing quantity to compute $k_L$ is the molar oxygen flux $\dot{N}_\Sigma$. If the concentration gradient at $\Sigma$ was known, the oxygen flux could be computed directly via Fick's law. Unfortunately, it is not possible to evaluate local concentration gradients in the experiment due to the small width of the concentration boundary layer and the strong light-reflections at the interface. Nevertheless, $\dot{N}_\Sigma$ can be determined via the temporal evolution of the oxygen concentration in the liquid bulk. Therefore, consider a time-dependent liquid volume $\Omega_l (t)$ enclosing the bubble as sketched in figure \ref{fig:domain_kl}.

The temporal evolution of $c_{O_2}(t,\mathbf{x})$ within $\Omega_l (t)$ may be computed using the Reynolds transport theorem:

\begin{equation}
\label{equ:reynolds_theorem}
    \frac{\mathrm{d}}{\mathrm{d}t} \int_{\Omega_l (t)} c_{O_2} \mathrm{d} V =
    \int_{\Omega_l (t)} \frac{\partial c_{O_2}}{\partial t}\mathrm{d} V +
    \int_{\partial\Omega_l (t)} c_{O_2} \mathbf{u}_b \cdot \mathbf{n}_b \mathrm{d} A\ ,
\end{equation}
where $\partial\Omega_l (t)$ is the boundary of $\Omega_l(t)$, $\mathbf{u}_b (t,\mathbf{x})$ is the velocity of $\partial\Omega_l (t)$, and $\mathbf{n}_b$ is an outward-pointing unit vector normal to $\partial\Omega_l (t)$. The concentration within $\Omega_l (t)$ is time-dependent and may change due to convection, diffusion, and chemical reactions. Therefore, the partial time derivative is to be replaced by

\begin{equation}
\label{equ:time_derivative}
    \frac{\partial c_{O_2}}{\partial t} = \nabla\cdot (D_{O_2}\nabla c_{O_2} - \mathbf{u}c_{O_2}) + \dot{r}\ ,
\end{equation}
where $\nabla$ denotes the Nabla operator, and $\mathbf{u}(t,\mathbf{x})$ is the velocity vector in the liquid phase. Substituting equation \eqref{equ:time_derivative} in \eqref{equ:reynolds_theorem} and applying the divergence theorem leads to

\begin{equation}
\label{equ:full_dt_equation}
    \frac{\mathrm{d}}{\mathrm{d}t} \int_{\Omega_l (t)} c_{O_2} \mathrm{d} V =
    \int_{\Omega_l (t)} \dot{r} \mathrm{d} V +
    \int_{\partial\Omega_l (t)}\left[ D_{O_2}\nabla c_{O_2} + c_{O_2} (\mathbf{u}_b-\mathbf{u})\right]\cdot \mathbf{n}_b \mathrm{d} A\ .
\end{equation}
If we consider a co-moving domain of fixed shape $\Omega_l$ and assume all processes to be steady, the temporal change of $c_{O_2}$ within $\Omega$ becomes zero.  Furthermore, the integral over the boundary $\partial \Omega_l$ may be decomposed into two parts: the gas-liquid interface $\partial\Omega_\Sigma$ and the outer boundary $\partial\Omega_o$. Simplifying and reordering equation \eqref{equ:full_dt_equation} leads to

\begin{align}
\label{equ:steady_equation}
    \int_{\Omega_l} \dot{r} \mathrm{d} V +
    \int_{\partial\Omega_\Sigma} \left[ D_{O_2}\nabla c_{O_2} - c_{O_2}(\mathbf{u}_b-\mathbf{u}) \right]\cdot \mathbf{n}_b \mathrm{d} A &+\notag\\
    \int_{\partial\Omega_o} \left[ D_{O_2}\nabla c_{O_2} - c_{O_2} (\mathbf{u}_b-\mathbf{u}) \right]\cdot \mathbf{n}_b \mathrm{d} A &= 0\ .
\end{align}
If $\Omega_l$ is co-moving with the bubble's center of mass, the velocity normal to $\partial \Omega_l$ is equal to the bubble rise velocity $\mathbf{u}_b\cdot\mathbf{n}_b = U_b$ on $\partial\Omega_\Sigma$. On the outer domain boundary $\partial\Omega_o$, the sign changes because normal and velocity vector point in the opposite direction: $\mathbf{u}_b\cdot\mathbf{n}_b = -U_b$.  Because the bubble shape is steady, the interface velocity must be equal to the bubble rise velocity such that $(\mathbf{u}_b-\mathbf{u})\cdot \mathbf{n}_b = 0$ on $\partial\Omega_\Sigma$. If the outer boundary if far from the bubble, the liquid velocity is at rest again such that $\mathbf{u}=\mathbf{0}$ on $\partial\Omega_o$. The integral over the diffusive flux normal to $\partial\Omega_\Sigma$ is the molar flux $\dot{N}_\Sigma$ required to compute the mass transfer coefficient. The diffusive flux over the outer boundary may be assumed to be negligible since the wake is closed and the oxygen concentration changes only mildly in streamwise direction; see figure \ref{fig:pseudocolorimage}. After applying all aforementioned simplifications, equation \eqref{equ:steady_equation} becomes

\begin{equation}
\label{equ:reduced_equ_reaction}
    \dot{N}_\Sigma = -\int_{\Omega_l} \dot{r} \mathrm{d} V - \int_{\partial\Omega_o} c_{O_2} U_b \mathrm{d} A\ .
\end{equation}
Note that the negative sign of the oxygen flux means that it enters the control volume $\Omega_l$ because of the outward-pointing normal vector. Expression \eqref{equ:reduced_equ_reaction} states that molar flux transferred from the gas to the liquid phase is equal to the sum of oxygen consumption within $\Omega_l$ due to chemical reactions and the oxygen leaving $\Omega_l$ over the outer boundary $\partial\Omega_o$ due to the motion of the boundary. The planar LIF measurement provides concentration values within a plane through $\Omega_l$, which can be used to compute the integrals in equation \eqref{equ:reduced_equ_reaction} under the assumption of axis-symmetry around the rise velocity vector of the bubble. However, the computation of the volume integral over the reactive term in \eqref{equ:reduced_equ_reaction} poses new challenges compared to previous works: a significant amount of oxygen will be consumed within the boundary layer, but the boundary layer cannot be captured accurately. Thus, it is unlikely to obtain meaningful reactive mass transfer coefficients based on the present experimental approach. For experiments without chemical reaction, the surface integral can be computed as demonstrated in \cite{Kuc12}; see figure 5 in the reference for a detailed sketch.

\begin{figure}[ht]
	\centering
    \includegraphics[width=0.75\textwidth]{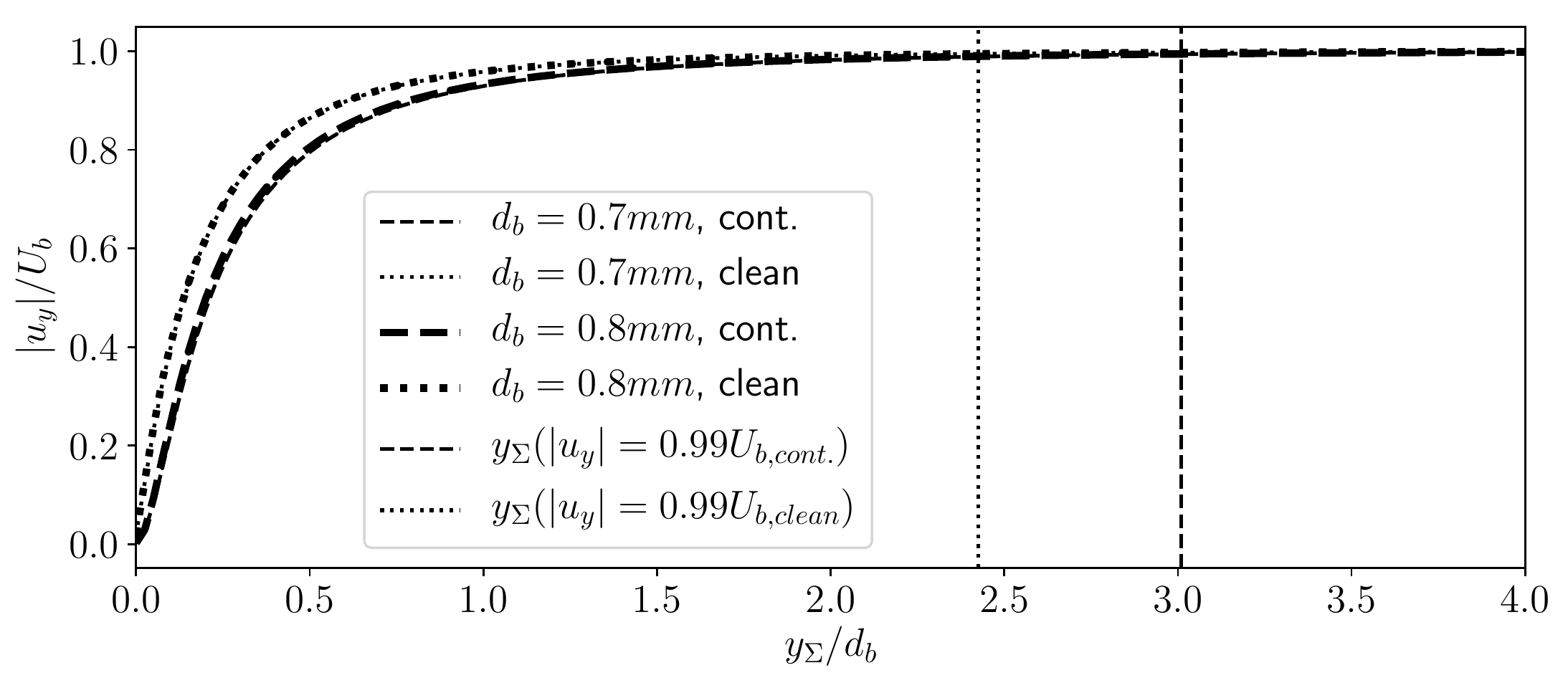}
    \caption{Ratio of the absolute velocity component in rise direction $|u_y|$ normalized with the bubble rise velocity $U_b$ plotted against the dimensionless distance to the gas-liquid interface $y_\Sigma / d_b$ on a center line (in rise direction, $y$) in the bubble wake. The distances $y_\Sigma(u_b=0.99U_{b,clean}) = 2.45d_b$ and $y_\Sigma(u_b=0.99U_{b,cont}) = 3.01d_b$ depict at which distance the liquid phase is close to the steady rise velocity in the moving reference frame or close to be at rest in a fixed frame of reference, respectively.}
	\label{fig:wake_center_velocity}
\end{figure}

Since the experimentally observable area is limited, it is of interest to know at what distance behind the bubble the perturbation of the velocity field has vanished. Therefore, we extracted the velocity component in rise-direction on the centerline in the bubble wake (numerical data). Figure \ref{fig:wake_center_velocity} shows the normalized velocity component plotted against the normalized distance from the bubble south-pole. At distances of $y_\Sigma = 2.45d_b$ for the clean and $y_\Sigma = 3.01d_b$ for the contaminated cases, the velocity field is perturbed by less than $1~\%$ relative to the rise velocity. The Marangoni forces introduce additional shear forces into the liquid bulk, which perturb the liquid bulk more than in the clean case. The length of the wake perturbation scales approximately linear with the bubble diameter for clean and contaminated cases.

\begin{figure}[ht]
	\centering
 	\includegraphics[width=0.75\textwidth]{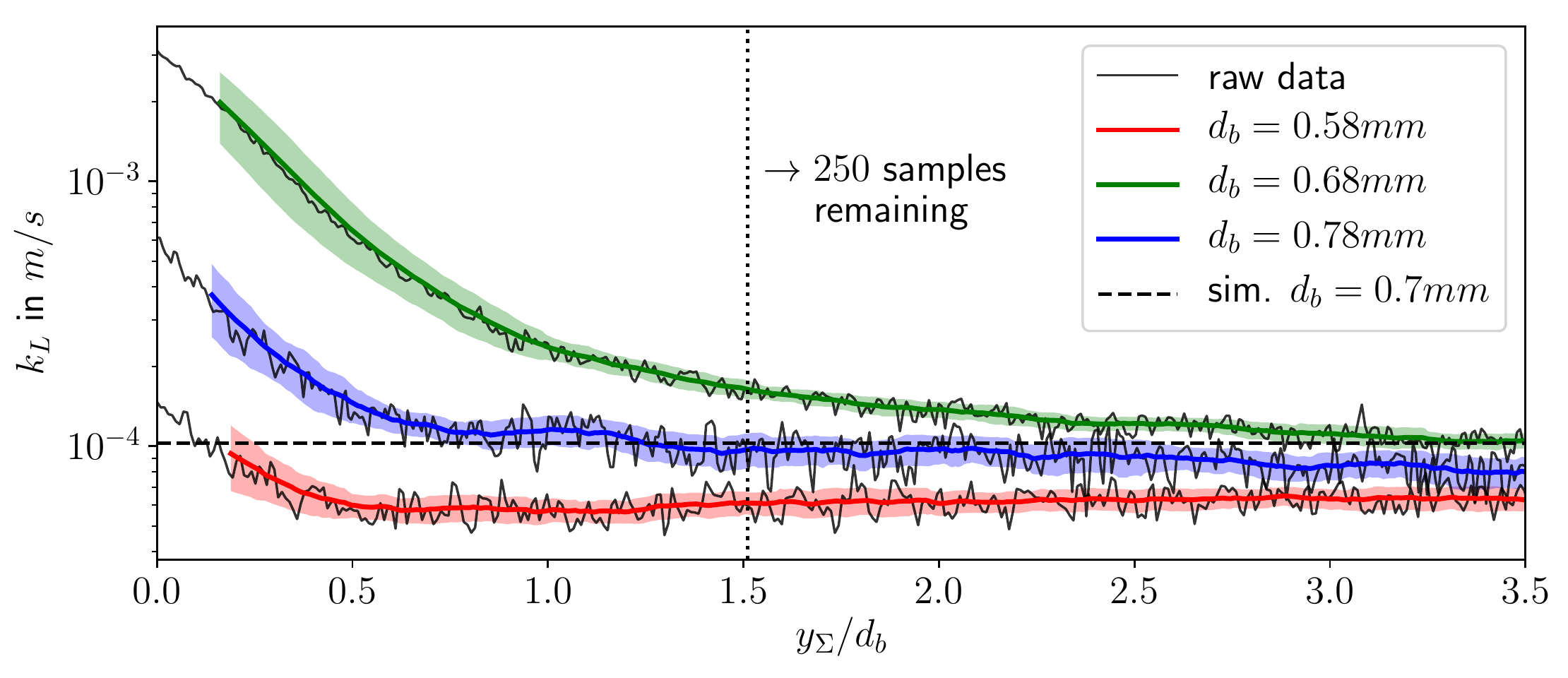}
	\caption{Computed mass transfer coefficient $k_L$ plotted against the normalized distance from the interface. The moving average and standard deviation (colored lines) were computed on a window size of $20$ samples. The shaded area around the moving average expresses the $68\%$ confidence interval ($\bar{k}_L\pm \sigma_{k_L}$).}
	\label{fig:kl_vs_distance}
\end{figure}

\begin{figure}[ht]
	\centering
 	\includegraphics[width=0.9\textwidth]{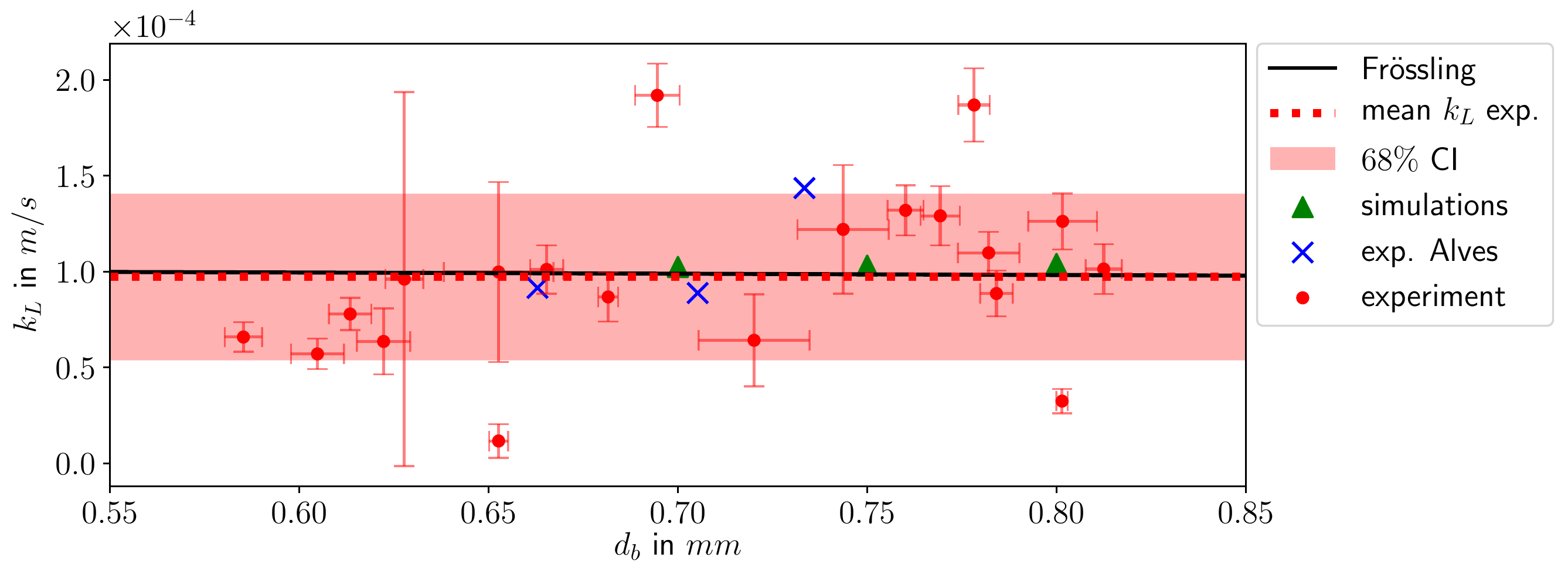}
	\caption{Mass transfer coefficient from experiments, simulations, and literature plotted against the bubble diameter. The reference line is computed based on the drag coefficient by Schiller and Naumann  (table \ref{tab:drag_coefficients}, solid particle) and the Sherwood number correlation by Fr\"ossiling (equation \eqref{equ:froessling}). Alvis experimental results are extracted from figure 10 in \cite{Alv05}.}
	\label{fig:kl_vs_diameter}
\end{figure}

Figure \ref{fig:kl_vs_distance} depicts, similar to figure \ref{fig:wake_center_velocity}, the dependency of the measured mass transfer coefficient on the distance from the bubble south-pole. After a distance of approximately $1.5d_b$, the $k_L$ signal becomes stable. Note that the mass transfer coefficient in the considered diameter range should be almost constant according to the literature references and the simulations. The dashed line in figure \ref{fig:kl_vs_distance} depicts an exemplary reference. However, the computed $k_L$ values are not monotonously changing with the diameter, and the deviation from the reference is much larger than the confidence interval. The standard deviation in the figures \ref{fig:kl_vs_distance} and \ref{fig:kl_vs_diameter} only reflects the changes in the measured concentration profiles. It is likely that the offset stems from errors in the measured diameter and rise velocity. The standard deviation of the rise velocity is larger than that of the diameter (see table \ref{tab:relative_std_ub_d}), but the error in $d_b$ influences the computed $k_L$ quadratically; see equation \eqref{equ:massflowrate5}. Furthermore, there is probably a systematic error in the measured concentration profile due to the finite thickness of the laser sheet, as discussed in section \ref{sec:wake_concentration}.

For the $k_L$ values depicted in figure \ref{fig:kl_vs_diameter}, the last $250$ lateral planes were averaged over the measured concentration profiles in the bubble wake. Because the mass transfer coefficient in the considered diameter range is almost constant, we also computed an average value with respect to the diameter (a linear trend line would yield a similar result). Despite the different error sources, the diameter-averaged mass transfer coefficient agrees very well with Fr\"ossling's correlation and also with the numerical predictions. The relative standard deviation of approximately $20~\%$ seems reasonable given the complexity of the overall measurement; see table \ref{tab:relative_std_kl} for the exact values. It is hard to reliably compare this value to other works because the overall confidence interval is usually not reported. 

\begin{table}[ht]
	\centering
	\caption{Minimum, maximum and mean relative standard deviation for the computed $k_L$.}
	\vspace{0.2cm}
	\renewcommand{\arraystretch}{1.3}
	\begin{tabular}{ccc}
		\toprule
		 $\mathrm{min}(\sigma_{k_L}/k_L)$ & $\mathrm{max}(\sigma_{k_L}/k_L)$ & $\mathrm{mean}(\sigma_{k_L}/k_L)$ \\
		\midrule
        $0.0860$ & $1.0146$ & $0.2437$ \\
		\bottomrule
	\end{tabular}
	\label{tab:relative_std_kl}
\end{table}

\begin{figure}[ht]
	\centering
 	\includegraphics[width=0.9\textwidth]{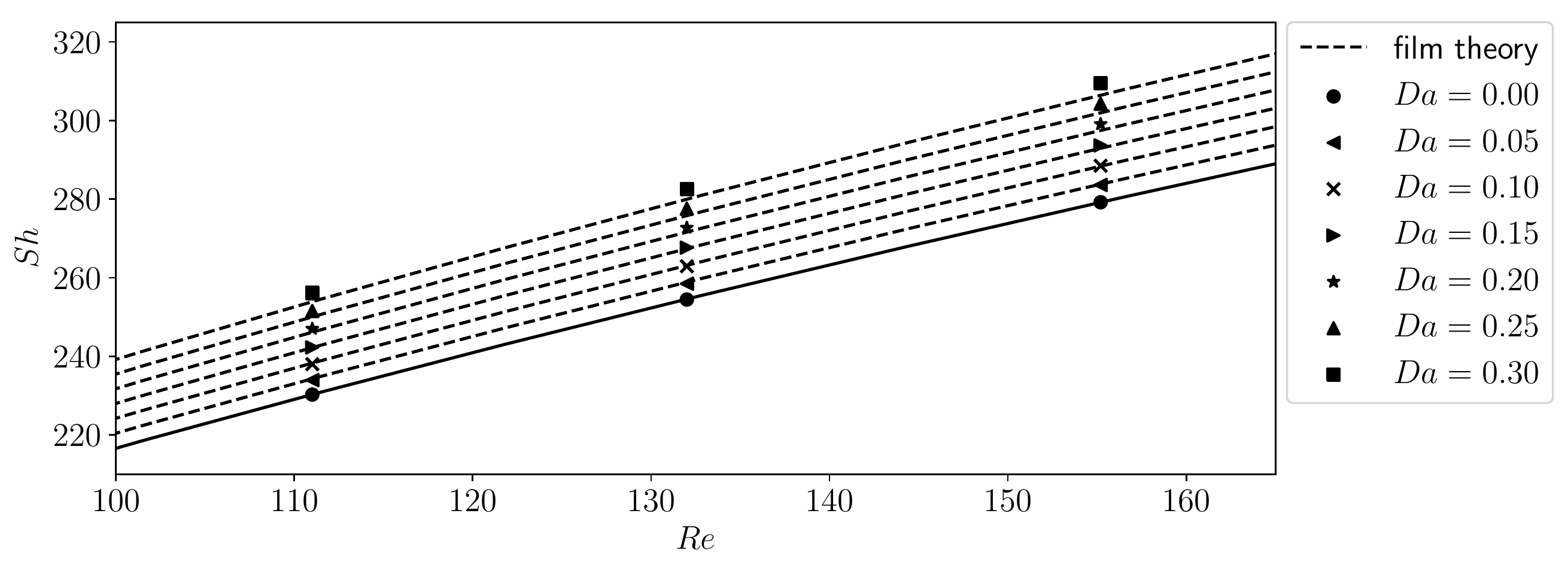}
	\caption{Reactive Sherwood number for different Damk\"ohler numbers plotted against the Reynolds number for a clean interface. The non-reactive $Sh$ value was computed based on the drag of Haas (table \ref{tab:drag_coefficients}) and the modified version of Lochiel's correlation for mass transfer \eqref{equ:modified_lochiel}.}
	\label{fig:Sh_reactive_clean}
\end{figure}

\begin{figure}[ht]
	\centering
 	\includegraphics[width=0.9\textwidth]{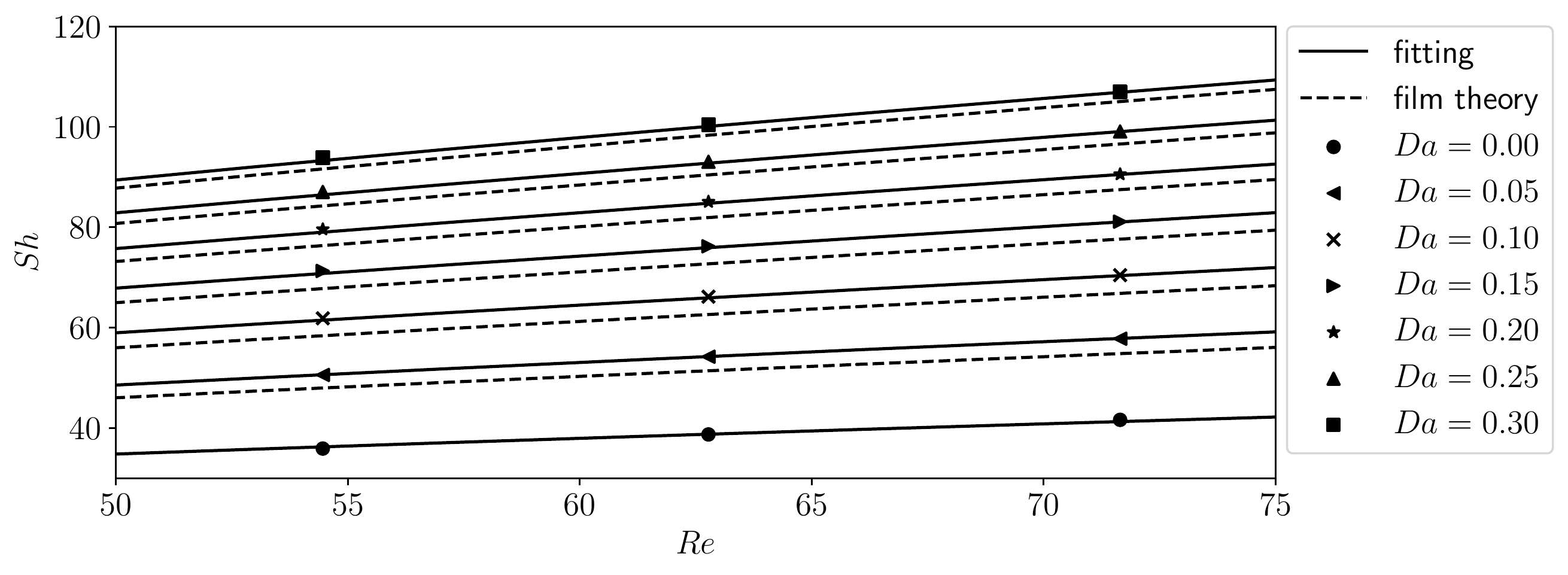}
	\caption{Reactive Sherwood number for different Damk\"ohler numbers plotted against the Reynolds number for a fully contaminated interface. The non-reactive $Sh$ value is computed based on the drag of Schiller and Naumann (table \ref{tab:drag_coefficients}) and a modified version of Fr\"ossling's correlation for mass transfer \eqref{equ:modified_froessling}. The \textit{fitting} corresponds to equation \eqref{equ:modified_enhancement}.}
	\label{fig:Sh_reactive_cont}
\end{figure}

In table \ref{tab:clean_kl} we also reported the numerical results for mass transfer at a clean bubble. A comparison with experimental data is not possible due to the inevitable presence of the fluorescence tracer. However, the deviation from different literature correlations is marginal. A modification of the coefficient in equation \eqref{equ:sh_lochiel} fits the numerical results even better:

\begin{equation}
    Sh = \frac{2}{\sqrt{\pi}}\left( 1-\frac{\mathbf{2.63}}{Re^{1/2}} \right)^{1/2}\sqrt{Pe}\ .
    \label{equ:modified_lochiel}
\end{equation}
Note that this modification is not meant as a ``correction'' but rather a step to avoid a propagation of the deviation into the reference curves for the reactive Sherwood numbers reported later on in figures \ref{fig:Sh_reactive_clean} and \ref{fig:Sh_reactive_cont}. 

\begin{table}[ht]
	\centering
	\caption{Predicted $k_L$/$Sh$ for a mobile interface: present simulations, correlation by Takemura and Yabe \cite{takemura1998} and correlation by Lochiel and Calderbank \cite{lochiel1964}. Both correlations can be found in \ref{sec:mass_transfer_correlations}. The Reynolds number to evaluate $Sh(Re, Sc)$ was computed based on the numerically predicted rise velocity.}
	\vspace{0.2cm}
	\renewcommand{\arraystretch}{1.3}
	\begin{tabular}{ccccccc}
		\toprule
		 $d_b$ in $mm$ & $Re$ & $Sc$ & $k_L$ in $m/s$ &$Sh$ & eq. \ref{equ:sh_takemura} & eq. \ref{equ:sh_lochiel}  \\
		\midrule
        $0.70$ & $111$ & $500$ & $3.29\cdot 10^{-4}$  & $230$ & $227$ & $225$ \\
        $0.75$ & $132$ & $500$ & $3.39\cdot 10^{-4}$  & $254$ & $250$ & $250$ \\
        $0.80$ & $155$ & $500$ & $3.49\cdot 10^{-4}$  & $279$ & $274$ & $274$ \\
		\bottomrule
	\end{tabular}
	\label{tab:clean_kl}
\end{table}

The numerically computed mass transfer coefficients for the contaminated cases agree very well with Fr\"ossling's correlation. It is remarkable that the sheer influence of Marangoni forces on the boundary layer structure leads to a more than $6$ times lower mass transfer coefficients/Sherwood numbers. A minor modification of equation \eqref{equ:froessling} that better fits the numerical prediction reads

\begin{equation}
    Sh = 2 + \mathbf{0.5847} Sc^{\frac{1}{3}} Re^{\frac{1}{2}}\ .
    \label{equ:modified_froessling}
\end{equation}

As we reported in section \ref{sec:wake_concentration}, already a slow chemical reaction leads to a substantial depletion of the oxygen concentration in the bubble wake. Because of the reason mentioned above, it is not possible to measure meaningful reactive mass transfer coefficients. However, we can provide insightful numerical results which we compare to the enhancement predicted by film-theory \eqref{equ:enhancement_film_theory}.

\begin{table}[ht]
	\centering
	\caption{Predicted $k_L$/$Sh$ for a (partially) immobile interface: present simulations and correlation by Fr\"ossling \cite{Fro38} (equation \ref{equ:froessling}). The Reynolds number to evaluate $Sh(Re, Sc)$ is computed based on the numerically predicted rise velocity.}
	\vspace{0.2cm}
	\renewcommand{\arraystretch}{1.0}
	\begin{tabular}{cccccc}
		\toprule
		 $d_b$ in $mm$ & $Re$ & $Sc$ & $k_L$ in $m/s$ &$Sh$ & eq. \ref{equ:froessling}  \\
		\midrule
        $0.70$ & $54$ & $500$ & $1.03\cdot 10^{-4}$  & $36$ & $34$ \\
        $0.75$ & $63$ & $500$ & $1.04\cdot 10^{-4}$  & $39$ & $37$ \\
        $0.80$ & $72$ & $500$ & $1.05\cdot 10^{-4}$  & $42$ & $39$ \\
		\bottomrule
	\end{tabular}
	\label{tab:cont_kl}
\end{table}

For the fully mobile interface the increase of the Sherwood number is small, $ E = Sh_{r}/Sh \le 1.1 $; see figure \ref{fig:Sh_reactive_clean} for the investigated reactive time scales. In this range, the relative difference between film theory and simulation is less than $1~\%$.

For the partially immobilized interface the reactive enhancement is much stronger $E \le 2.6$; see figure \ref{fig:Sh_reactive_cont}. We investigated the same ratio of convective to reactive timescale $Da$, meaning that we decreased the reaction rate corresponding to the decrease in the rise velocity. This adjustment is necessary to isolate the effect of surface contamination on mass transfer. In an experiment comparing clean and contaminated bubbles one would probably keep the reaction rate constant, such that $Da$ increases significantly for the slower rising, contaminated bubbles. The observed reactive enhancement would be therefore even stronger than reported here.

The film theory predictions deviate up to $6~\%$ from the numerical results. An empirical modification of equation \eqref{equ:enhancement_film_theory} that fits the numerical results within a $1~\%$ range reads

\begin{equation}
    E_{cont} = \frac{f(Ha)}{\mathrm{tanh}\left( f(Ha) \right)}\quad \text{with} \quad f(Ha) = 1.117 Ha^{0.9} \ .
    \label{equ:modified_enhancement}
\end{equation}

\begin{figure}[ht]
	\centering
 	\includegraphics[width=0.9\textwidth]{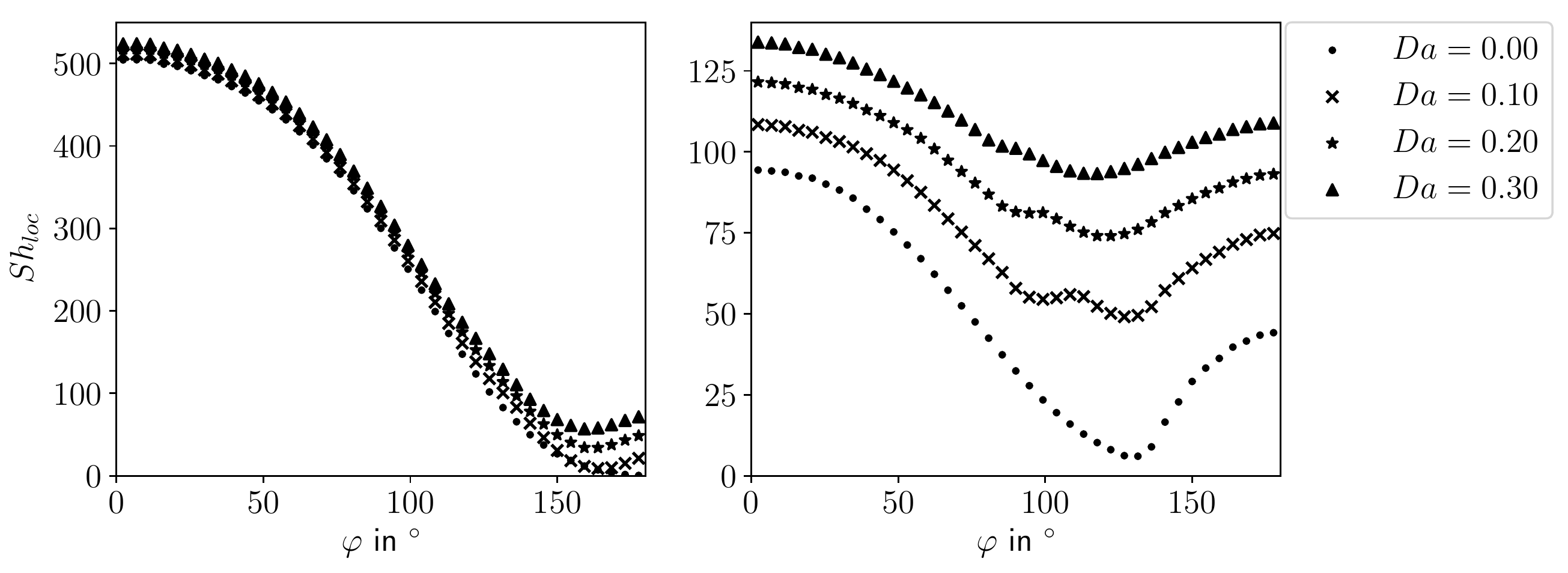}
	\caption{Local reactive Sherwood number plotted against the polar angle $\varphi$ for a clean (left) and contaminated (right) interface.}
	\label{fig:local_sh}
\end{figure}

Finally, we would like to explain why the film theory prediction of the enhancement factor for clean bubbles is more accurate than for contaminated ones. Figure \ref{fig:local_sh} shows the local Sherwood number plotted over the polar angle. As already observed for the integral $Sh$ values, the enhancement at the immobilized interface is much stronger. The main difference, however, results from the non-uniform local mass transfer enhancement in the presence of surfactant. Especially in the rear part of the bubble, where the contact time between fluid-elements and interface is the largest, the enhancement is stronger than at the upper bubble half. Agreement with the film theory can only be achieved when the enhancement is uniform and small, like for the fully mobile interface.

\subsection{Discussion of surface contamination}
\label{sec:surface_contamination}
  The addition of the fluorescent tracer particles caused a significant drop in the observed rise velocities. The influence of a surfactant was already observed by K\"uck et al. \cite{Kuc10, Kuc11, Kuc12}, but the source could not be isolated. Jimenz et al. \cite{Jim14} investigated the impact of (intentionally added) surfactants on mass transfer at rising bubbles. The authors of \cite{Jim14} also measured the influence of the same fluorescent dye (by the same provider) as used in this work on several system properties; see table 1 on page 113 in the reference. While there is no measurable influence on the water's density or viscosity at room temperature, the surface tension was reduced by approximately $0.5~\%$. This marginal change seems to be negligible at first sight. However, the difference of the flow around a rising bubble and the reduction of the rise velocity in a contaminated system mostly stem from Marangoni-forces, which are equal to the surface tension gradient and not to the surface tension itself. Simply said, a small change of surface tension that occurs over a small surface region will have a similar effect as a larger reduction happening over a larger region. Thus, in the surface tension dominated regime, a marginal change of the surface tension may result in a substantial difference of the macroscopic behavior. In fact, the difference between the maximum and minimum surface tension in the numerical simulation is only about $0.9~\%$. This change of the jump-conditions at the interface leads to a rise velocity reduction of about $50~\%$.
Despite the evidence presented before, we cannot assure that the fluorescent dye is the surface active agent. There are several issues which prevent us from drawing clear conclusions:

\begin{itemize}
  \item Tap water was used in the experiment. Because tap water naturally contains contaminants, there could be the effect of several surfactants acting at the same time. The difference in the recorded rise velocities with and without fluorescent tracer could also stem from different degrees of contamination in the tap water since the experiments were not conducted on the same day.
  \item The surface tension reduction, measured by Jimenz et al. \cite{Jim14}, is very small and could have been below the accuracy range of the measurement device.
  \item The purity of the fluorescent dye is only $98~\%$. The additional impurities could alter the surface properties, too\footnote{We contacted the provider of the dye, but a potential surface activity could be neither confirmed nor refuted.}. Possible impurities are ruthenium chloride ($RuCl_3$) and biby (2,2´-Bipyridine\footnote{https://en.wikipedia.org/wiki/2,2\%27-Bipyridine}) which are left over from the synthesis of the fluorescent tracer. In water, the ruthenium chloride will dissociate into ruthenium and chloride ions ($Ru^{2+}$, $Cl^-$), which are not surface active. The bipy, however, is hydrophobic and could adsorb onto the interface.
\end{itemize}
Despite the ambiguous arguments presented before, it is not unlikely that the fluorescent dye caused the drastic drop of the rise velocity in the experiments. Therefore we discuss its molecular structure briefly hereafter.

\begin{figure}[ht]
    \centering
    \includegraphics[width=0.6\textwidth]{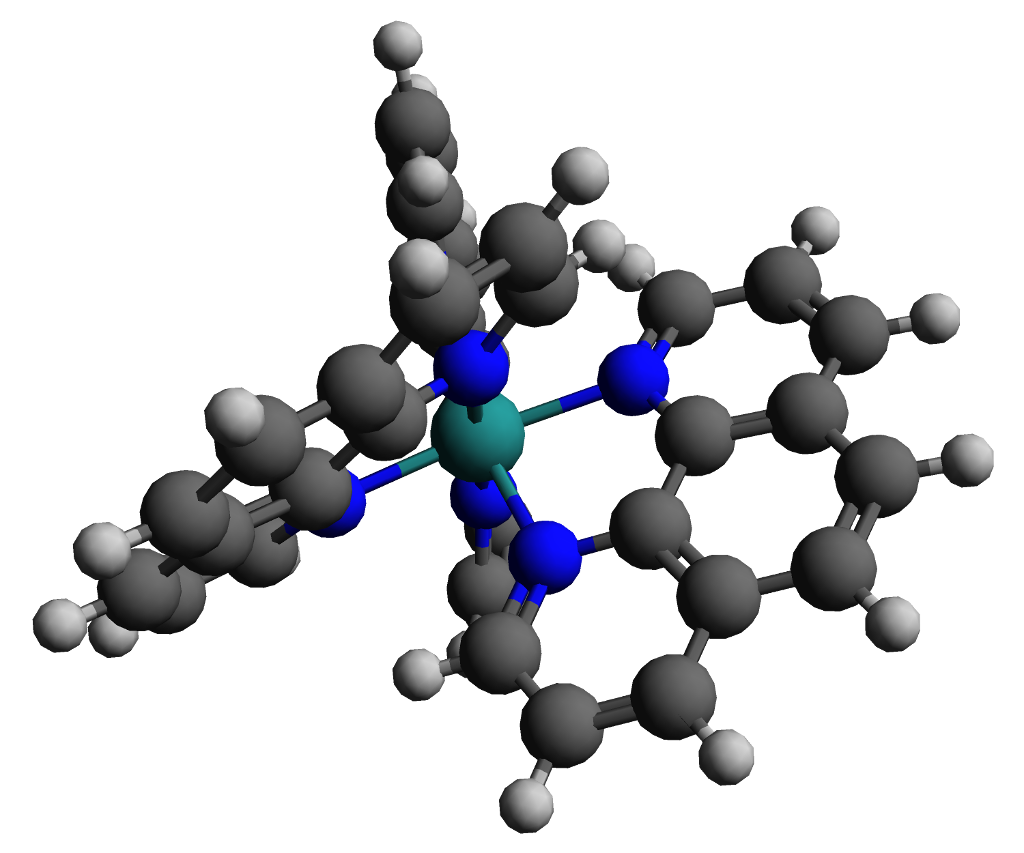}
    \caption{Fluorescent tracer: 3D model of the ruthenium complex cation \cite{turel2015}.}
    \label{fig:ruth_molecule}
\end{figure}

Figure \ref{fig:ruth_molecule} depicts the spacial structure of the fluorescent dye (crystalline structure). The central ruthenium atom (turquoise) is twice positively charged. This charge is distributed by the organic phenanthroline ligands surrounding the central atom (blue - nitrogen, dark gray - carbon, light gray - hydrogen). Somewhere in the space around this molecule, there are two negatively charged chloride ions (not depicted) to complete the charge balance. Both the ruthenium complex cation as well the chloride ions are charged, but the charge density of the chloride is much larger since the atoms are much smaller. Moreover, the organic ligand has hydrophobic properties. Compared to the ``classic'' structure of surfactants consisting of a ``polar'' head (hydrophilic) and a non-polar ``tail'' (hydrophobic), the fluorescence tracer could behave like a hydrophobic sphere (the ruthenium complex cation) with one or two polar heads (the chloride ions). Thus, it could potentially be surface active. To draw definite conclusions, further experiments under well-defined conditions are necessary. A very sensitive way to investigate the adsorption of a chemical species onto the interface is the measurement of the bubble rise velocity over time (so-called local velocity profiles \cite{pesciSPP2017}). Under the influence of soluble surfactant, the rise velocity will show three very distinct stages of acceleration, deceleration, and steady state \cite{Pesci2017}. Of course, such experiments would have to be carried out using ultra-purified water and a fluorescent tracer with higher purity.

\FloatBarrier

\section{Conclusion}
\label{sec:conclusion}
  In this work, we investigated the reactive mass transfer from small rising oxygen bubbles by means of experiments and numerical simulations. The high-speed planar-LIF technique is an excellent tool to determine the mass transfer coefficient of such small bubbles. Based on the averaging procedure of several ROI-images, the signal to noise ratio could be improved significantly, such that a more accurate determination in comparison with \cite{Kuc10, Kuc11, Kuc12} was possible. The fluorescent dye used in the experiments led to a strong reduction of the observed rise-velocity as it is typical for surfactants. We therefore discussed the potential polarity of the ruthenium complex and included the transport and surface tension reduction due to a surfactant in the numerical method. The simulations proved to have reached a high level of accuracy which allows to determine the bubble hydrodynamics and mass transfer for realistic time- and length-scales. Thus, simulations can validate experimental results and vice-versa. They also provide valuable insights into local quantities close to the interface.

There are several possible improvements to increase the quality of the experimental and numerical techniques further. In the experiments, a reduction of the uncertainty in each quantity involved in the calculation (diameter, rise velocity, wake concentration) will ultimately lead to a narrower confidence interval for the mass transfer coefficient. Future work will also be focused on employing novel reaction systems, such that educts and products can be detected simultaneously. On the simulation side, the subgrid-scale modeling has to undergo further developments to handle coupled higher-order reactions. Also, a mesh-topology with more control-volumes concentrated in the bubble wake would yield sharper concentration fields at larger distances from the interface.

\section*{Acknowledgement}

The authors gratefully acknowledge the financial support provided by the German Research Foundation within the Priority Program “Reactive Bubbly Flows”, SPP 1740 (SCHL 617/12-1 and BO 1879/13-2). The molecular structure of the ruthenium-complex and its interpretation were provided by the group of Prof.~S.~Schindler, a SPP 1740 project partner (SCHI 377/13-1). We would also like to thank Prof.~S.~Herres-Pawlis (HE 5480/10-2) and Dr. habil.~R.~Miller for fruitful discussions, which significantly helped to improve the manuscript. Numerical simulations were performed on the Lichtenberg high performance computer of the TU Darmstadt.

\appendix

\section{Additional resources}
\label{sec:appendix}
  \subsection{Correlations for the drag coefficient}
\label{sec:drag_correlations}

The literature contains many drag correlations for spherical particles at higher Reynolds number. The relationships are typically based on theoretical analysis, simulations or experiments. Table \ref{tab:drag_coefficients} lists some of the most common correlations, which we used to verify our experimental and numerical data. The correlations by Tomiyama et al. \cite{Tom98} are mainly hybrids of simpler relations in order to extend their range of validity. In the range of Reynolds numbers considered in the present work, the correlation by Tomiyama et al. for clean bubbles is equivalent to Levich's potential flow analysis, and the relation for contaminated bubbles is identical to Schiller's drag coefficient for solid particles. Note also that ``clean'' in experimental investigations means that no surface active agent was added intentionally. However, it is well known that the terminal velocity of bubbles rising in tap water may be reduced by up to $50~\%$ compared to bubbles rising in purified (deionized) water (see for example figure 7.3 on page 172 in \cite{clift1978}).

\begin{table}[ht]
	\centering
	\caption{Correlations for drag coefficients in pure and contaminated media from \cite{clift1978,Tom98}.}
	\vspace{0.2cm}
	\renewcommand{\arraystretch}{1.3}
	\begin{tabular}{lll}
		\toprule
		first author/source & system & drag $c_D$ \\
		\midrule
		Levich, eq. 5-30 in \cite{clift1978} & clean &
		$48/Re$\\
		Moore, eq. 5-31 in \cite{clift1978} & clean &
		$\frac{48}{Re}\left[ 1 - \frac{2.21}{Re^{1/2}} \right] $ \\
		Haas, eq. 5-28 in \cite{clift1978} & clean &
		$14.9Re^{-0.78}$ \\
		Schiller, tab. 5.1 in \cite{clift1978} & solid &
		$\frac{24}{Re}\left( 1 + 0.15 Re^{0.687} \right)$ \\
		Tomiyama, eq. 10 in \cite{Tom98} & clean &
		$\textrm{min}\left[\frac{16}{Re}\left(1 + 0.15Re^{0.687}\right),\frac{48}{Re}\right]$\\
		Tomiyama, eq. 13 in \cite{Tom98} & slightly cont. &
		$\textrm{min}\left[\frac{24}{Re}\left(1 + 0.15Re^{0.687}\right),\frac{72}{Re}\right]$\\
		Tomiyama, eq. 11 in \cite{Tom98} & fully cont. &
		$\frac{24}{Re}\left( 1 + 0.15 Re^{0.687} \right)$ \\
		\bottomrule
	\end{tabular}
	\label{tab:drag_coefficients}
\end{table}

\subsection{Correlations for the non-reactive mass transfer}
\label{sec:mass_transfer_correlations}

Lochiel and Calderbank \cite{lochiel1964} derived an analytical expression for the Sherwood number at high Schmidt and Reynolds numbers based on boundary layer theory. The simplified equation for bubbles reads (equation 86 in the reference)

\begin{equation}
\label{equ:sh_lochiel}
  Sh = \frac{2}{\sqrt{\pi}}\left( 1-\frac{2.96}{Re^{1/2}} \right)^{1/2}\sqrt{Pe}\ .
\end{equation}

Takemura and Yabe \cite{takemura1998} modified relation \eqref{equ:sh_lochiel} based on their experimental data in order to cover a broader range of Reynolds numbers. Note that relation (22) in the reference is miss-printed. The correct correlation reads

\begin{equation}
\label{equ:sh_takemura}
  Sh = \frac{2}{\sqrt{\pi}}\left( 1-\frac{2}{3 \left(1+0.09Re^{2/3}\right)^{3/4}} \right)^{1/2}(2.5+\sqrt{Pe})\ .
\end{equation}

\subsection{Reactive subgrid-scale model}
\label{sec:reactive_sgs_model}

\begin{figure}[ht]
    \centering
    \begin{tikzpicture}
    \coordinate (y) at (0,-5);
    \coordinate (x) at (5,0);
    \draw[<->] (y) node[below] {$y$} -- (0,0) --  (x) node[right] {$x$};

   \coordinate (A) at (0,-1);
   \coordinate (B) at (1.5,-4);
   \coordinate (C) at (4,-4);
   \draw[black,thick] (A) to[out=-80,in=180] (B) -- (C);
   \node[fill=black,circle, minimum size = 0.15cm, inner sep=0pt] at (A) {};
   \draw[] (A) node[left] {$c|_{\scriptscriptstyle \Sigma}$};
   \draw[] (C) node[above] {$c(x,y)$};

   \coordinate (C) at (1.5,-4.5);
   \draw[scale=0.5,domain=0:9,smooth,variable=\y,black,rotate=-90,samples=300,densely dotted,thick] plot(\y,{sqrt(\y)});
   \draw[](C) node[below]{$\delta\left(y\right)$};

   \coordinate (s1) at (1.5,-0.5);
   \coordinate (s2) at (2.5,-0.5);
   \coordinate (s3) at (3.5,-0.5);
   \coordinate (e1) at (1.5,-1.5);
   \coordinate (e2) at (2.5,-1.5);
   \coordinate (e3) at (3.5,-1.5);
   \draw[->,thick] (s1)--(e1);
   \draw[->,thick] (s2)--(e2);
   \draw[->,thick] (s3)--(e3);
   \draw[] (e3) node[below right] {$\mathbf{u}=\left(0,v\right)$};

    myball/.style={shade, ball color=black, circular drop shadow={
    shadow xshift=0pt, shadow yshift=0pt}};
   \coordinate (bc) at (-3,-2.5);
   \shadedraw[shading=ball,ball color=black!20, white] (bc) circle (1);

    \draw[->] (-3,0) to [out=-90,in=150](-2.6,-1.5);
    \draw[->] (-2.6,-1.5) to[out=-30,in=90] (-1.95,-2.5);
    \draw[->] (-1.95,-2.5) to[out=-90,in=30] (-2.6,-3.5);
    \draw[->] (-2.6,-3.5) to[out=210,in=90] (-3,-5.0);

   \coordinate (rad) at (-2,-2.5);
   \node[draw=black, minimum size=0.5cm, circle,thick] at (rad) {};
   \coordinate (up) at (-2,-2.25);
   \coordinate (down) at (-2,-2.75);
   \draw[dashed] (up) -- (0,0);
   \draw[dashed] (down) -- (0,-4.75);

   \coordinate (sig) at (-0.5,-3);
   \draw[inner sep=0.1em] (sig) node[above left] {$\Sigma$};
   \draw[] (sig) -- (0,-3.5);
   \node[] at (-1,-0.5) {$\Omega^-$};
   \node[] at (1,-0.5) {$\Omega^+$};

\end{tikzpicture}
    \caption{Simplified 2D model for species transport close to the bubble surface, figure based on \cite{Weiner2017}.}
    \label{fig:sgs_setting}
\end{figure}
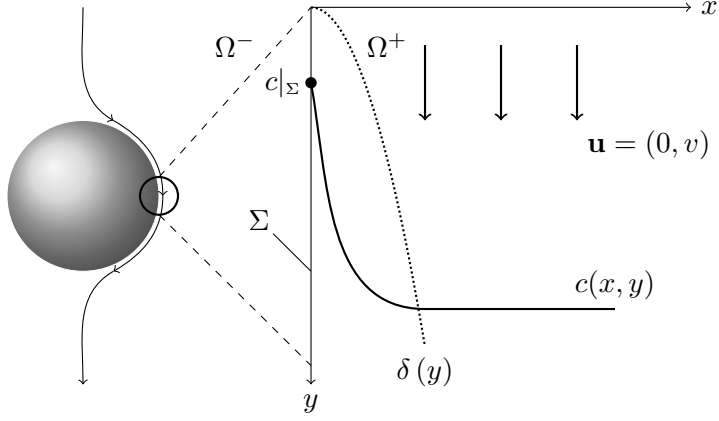

For this work, we extended the non-reactive subgrid-scale model \cite{Pesci2017} to handle convection-dominated species transfer accompanied by a first-order decay reaction. The non-reactive model was first introduced in \cite{Weiner2017} and then included in the interface-tracking framework \cite{Pesci2017}. A first step towards a reactive model was performed by Gr\"unding et al. \cite{gruending2016}. Here we only outline the main steps of the extension. A detailed description of the basic ideas and algorithms can be found in the cited literature. The substitute setting sketched in figure \ref{fig:sgs_setting} for the decay reaction is the same as for the non-reactive model. The initial boundary value problem describing the reactive boundary layer consists of the partial differential equation

\begin{equation}
\label{equ:reactive_pde}
  v\frac{\partial c}{\partial y} = D \frac{\partial^2 c}{\partial x^2} - kc\ ,
\end{equation}
accompanied by the initial and boundary conditions

\begin{equation}
\label{equ:initial_boundary_values}
  c(x>0,y=0) = 0;\quad c(x=0,y) = c_\Sigma;\quad c(x\rightarrow\infty, y) = 0\ .
\end{equation}
The constants appearing in equation \eqref{equ:reactive_pde} are $v$ - the tangential velocity component, $D$ - the molecular diffusivity, and $k$ - the reaction rate constant.

The solution for problem \eqref{equ:reactive_pde}/\eqref{equ:initial_boundary_values} may be computed using Laplace transformation and reads
\begin{equation}
\label{equ:pde_solution}
  \bar{c}(x,\widetilde{y}) = \frac{c}{c_\Sigma} 
 =\frac{e^{xA}}{2} \mathrm{erfc}\left( \frac{xC}{\widetilde{y}} +\widetilde{y}B\right)
 +\frac{e^{-xA}}{2} \mathrm{erfc}\left( \frac{xC}{\widetilde{y}} -\widetilde{y}B\right)\ ,
\end{equation}
with the abbreviations

\begin{equation}
\label{equ:symbol_groups}
  \widetilde{y} = \sqrt{y};\quad A=\sqrt{\frac{k}{D}};\quad B = \sqrt{\frac{k}{v}};\quad C = \sqrt{\frac{v}{4D}}\ .
\end{equation}
In equation \eqref{equ:pde_solution}, $\mathrm{erfc}$ denotes the complementary error function.

\begin{figure}[ht]
    \centering
    \includegraphics[width=0.8\textwidth]{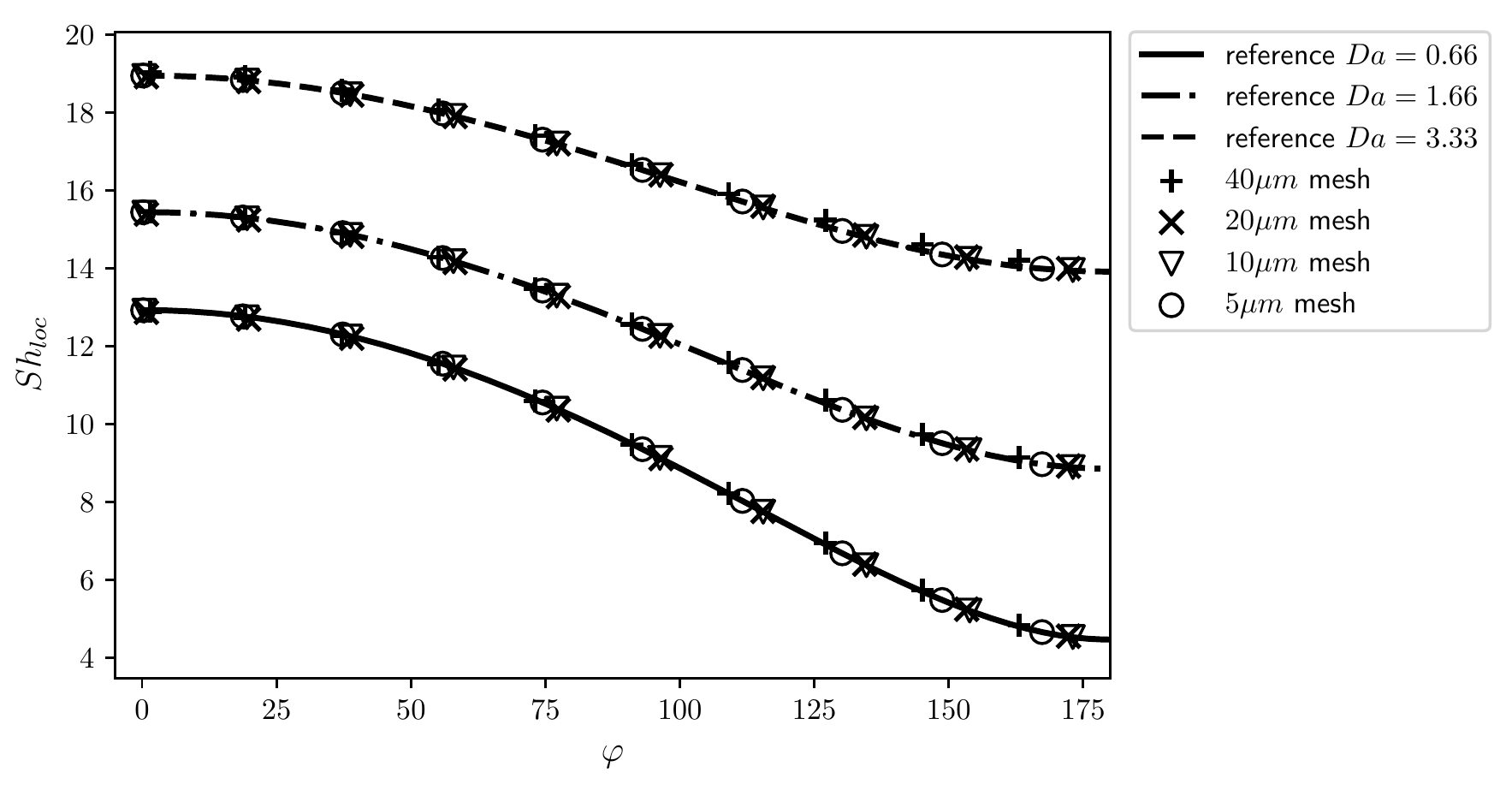}
    \caption{Local Sherwood number plotted against the polar angle for different reactive time scales and mesh resolutions. The P\'eclet number is $Pe=280$.}
    \label{fig:local_sh_validation}
\end{figure}

The next step in the modeling process is to adjust the analytical solution to the cell-centered concentration $\bar{c}_{num} = c_{num}/c_\Sigma$ of each interface cell in the numerical simulation. More specifically, the condition to compute the model parameter $\widetilde{y}$ reads

\begin{equation}
\label{equ:integral_fitting}
  \bar{c}_{num} - \frac{1}{L}\int\limits_{x=0}^L \bar{c}(x,\widetilde{y}) \mathrm{d}x = 0\ ,
\end{equation}
where $L$ is the thickness of the interface cell. An efficient iterative algorithm to find $\widetilde{y}$ is a Newton method based on equation \eqref{equ:integral_fitting}. The following two expressions are necessary for the implementation:

\begin{align}
  \int\limits_{x=0}^L \bar{c}(x,\widetilde{y}) \mathrm{d}x
  &= \frac{e^{LA}}{2A}
     \left[
     \mathrm{erfc}\left( \frac{LC}{\widetilde{y}} + \widetilde{y}B\right) +
     e^{-\left( \widetilde{y}B \right)^2-LA} \mathrm{erf}\left( \frac{LC}{\widetilde{y}} \right)
     \right]\label{equ:integral_over_x} \notag\\
  &- \frac{e^{-LA}}{2A}
     \left[
     \mathrm{erfc}\left( \frac{LC}{\widetilde{y}} - \widetilde{y}B\right) +
     e^{-\left( \widetilde{y}B \right)^2+LA} \mathrm{erf}\left( \frac{LC}{\widetilde{y}} \right)
     \right] \notag\\
  &+ \frac{\mathrm{erf}\left( \widetilde{y}B \right)}{A}\ ,\\
  \frac{\mathrm{d}}{\mathrm{d}\widetilde{y}}\int\limits_{x=0}^L \bar{c}(x,\widetilde{y}) \mathrm{d}x
  &= \frac{e^{-\left( \widetilde{y}B \right)^2}}{\sqrt{\pi}C}
     \left(
       1-e^{-\left( LC/\widetilde{y} \right)^2}
     \right)\ .
\end{align}
After $\widetilde{y}$ has been computed, convective and diffusive fluxes in the interface cells can be corrected. Equation \eqref{equ:pde_solution} can be used directly to get a more accurate approximation of the concentration value at $x=L$. For the diffusive fluxes at $x=0$ and $x=L$ two more relations have to be used:

\begin{align}
  \left.\frac{\partial \bar{c}}{\partial x}\right\vert_{x=0} &=
  -A\left[ \frac{e^{-\left( \widetilde{y}B \right)^2}}{\sqrt{\pi}\widetilde{y}B} + \mathrm{erf}\left( \widetilde{y}B \right)\right]\ ,\\
  \left. \frac{\partial \bar{c}}{\partial x}\right\vert_{x=L}
  &= \frac{A}{2}
     \left[
       e^{LA}\mathrm{erfc}\left( \frac{LC}{\widetilde{y}} + \widetilde{y}B \right)
     - e^{-LA}\mathrm{erfc}\left( \frac{LC}{\widetilde{y}} - \widetilde{y}B \right)
     \right]\notag\\
  &- \frac{2Ce^{-\left( LC/\widetilde{y} \right)^2-\left( \widetilde{y}B \right)^2}}{\sqrt{\pi}\widetilde{y}}\ .
\end{align}
The chemical reaction introduces another difficulty compared to the non-reactive model. As $\widetilde{y}$ becomes larger, the boundary layer and its thickness approach a steady state. In the limit $\widetilde{y}\rightarrow\infty$ the solution of the simpler film-theory is obtained:

\begin{equation}
\label{equ:limit_solution}
  \lim_{\widetilde{y}\rightarrow \infty} \bar{c}(x,\widetilde{y}) = e^{-xA}\ .
\end{equation}
Consequently, also the average model concentration value in relation \eqref{equ:integral_fitting} is limited by

\begin{equation}
\label{equ:average_x_limit}
  \bar{c}_{max} = \lim_{\widetilde{y}\rightarrow\infty} \frac{1}{L}\int\limits_{x=0}^L \bar{c}(x,\widetilde{y}) \mathrm{d}x =
  \frac{1-e^{-LA}}{LA}\ .
\end{equation}
If the cell-centered concentration value $\bar{c}_{num}$ is larger than expression \eqref{equ:average_x_limit}, it becomes impossible to find a suitable $\widetilde{y}$. Such values typically appear in the rear-part of a rising bubble. The normal velocity component close to the interface causes an additional convective transport of the chemical species, and hence the boundary layer becomes thicker. Because normal velocity contributions are not included in the substitute problem, the model function cannot be used without modification. A heuristic solution which works excellently consists of the following two steps:

\begin{enumerate}
  \item When the maximum average concentration $\bar{c}_{max}$ is smaller than $\bar{c}_{num}$ we fit the film theory solution by adjusting the parameter $A$. Decreasing $A$ corresponds to increasing the ratio of diffusive flux into the boundary layer to reactive decay. This adjustment emulates the additional convective flux normal to the interface in the numerical simulation.
  \item We do not apply the full SGS-model fluxes but rather a combination between standard (linearly interpolated) and SGS-model fluxes. A simple blending of the form $\phi = (1-w)\phi_{SGS} + w\phi_{num}$ works well with $\phi$ representing either the concentration value or its gradient. A sensible weight is $w=\bar{c}_{num}$. As $c_{num}$ approaches the interface value $c\vert_{\Sigma}$, the boundary layer is typically well resolved and no model correction is required. On the other hand, when $c_{num}$ is very small, a simple linear approximation will overestimate fluxes leaving the interface cell, while the SGS-model fluxes yield much better results.
\end{enumerate}
A standard test case for SGS-model validation is the semi-analytical solution of species transfer from a spherical bubble rising at a very low Reynolds number. The solution is termed semi-analytical because the species transport equation is solved numerically using an analytical expression for the velocity field to compute convective fluxes. Our main quality indicator is the local Sherwood number depicted over the polar angle. Figure \ref{fig:local_sh_validation} highlights two points: (i) The species transfer enhancement due to the chemical reaction in the boundary layer is captured well, and (ii) the solution shows very little dependence on the computational mesh. The reference solution was computed on a mesh where the first cell layer thickness was approximately $0.1~\mu m$. Even the global Sherwood numbers computed on the coarsest mesh, employing the subgrid-scale model, deviate by far less than $1~\%$ from the reference. We also repeated the validation procedure for $Pe$ numbers up to $Pe=10^6$, but the results were similar and are therefore not included here.

\FloatBarrier

\bibliographystyle{plain} 
\bibliography{references}

\end{document}